\documentclass[fleqn,usenatbib]{mnras}

\usepackage{newtxtext,newtxmath}

\usepackage[T1]{fontenc}
\usepackage{ae,aecompl}

\usepackage{graphicx}	
\usepackage{amsmath}	
\usepackage{booktabs}
\graphicspath{{images/}}

\usepackage{xspace}

\newcommand{\Msun}{\,{\rm M}_\odot}
\newcommand{\Mstar}{\,{\rm M}_*}

\newcommand{\JWST}{\textit{JWST}}

\title[Milky Way progenitors]{Observing EAGLE galaxies with \JWST: predictions for Milky~Way progenitors and their building blocks}

\author[T. A. Evans et al.]{
Tilly A. Evans$^{1,2}$,\thanks{E-mail: tilly.evans@durham.ac.uk}
Azadeh Fattahi$^{1}$,
Alis J. Deason$^{1,2}$
and Carlos S. Frenk$^{1}$
\\
$^{1}$Institute for Computational Cosmology, Department of Physics, Durham University, South Road, Durham, DH1 3LE, UK \\
$^{2}$Centre for Extragalactic Astronomy, Department of Physics, Durham University, South Road, Durham, DH1 3LE, UK\\
}

\date{Accepted XXX. Received YYY; in original form ZZZ}

\pubyear{2022}

\begin{document} 
\label{firstpage}
\pagerange{\pageref{firstpage}--\pageref{lastpage}}
\maketitle

\begin{abstract}
  We present predictions, derived from the EAGLE $\Lambda$CDM
  cosmological hydrodynamical simulations, for the abundance and
  properties of galaxies expected to be detected at high redshift by
  the {\it James Webb Space Telescope} (\JWST). We consider the galaxy
  population as a whole and focus on the sub-population of progenitors of Milky Way (MW) analogues, defined to be galaxies with accretion histories
  similar to the MW's, that is, galaxies that underwent a merger
  resembling the Gaia-Enceladus-Sausage (GES) event and that contain
  an analogue of the Large Magellanic Cloud (LMC) satellite today. We
  derive the luminosity function of all EAGLE galaxies in \JWST/NIRCam passbands, in the redshift range $z=2-8$, taking
  into account dust obscuration and different exposure times. For an
  exposure time of $T=10^5$s, average MW progenitors are observable as far
  back as $z\sim6$ in most bands, and this changes to $z\sim5$ and $z\sim4$ for the GES and LMC progenitors, respectively. The progenitors of GES and LMC
  analogues are, on average, $\sim 2$ and $\sim 1$ mag fainter
  than the MW progenitors at most redshifts. They lie, on average,
  within $\sim 60$ and $30$ arcsec, respectively, of their future MW
  host at all times, and thus will appear within the field-of-view of
  \JWST/NIRCam. We conclude that galaxies resembling the main progenitor of the MW and
  its major accreted components should be observable with \JWST\ beyond redshift $2$, providing a new and unique window in studying the formation history of our own galaxy.
\end{abstract}




\section{Introduction}

The {\it James Webb Space Telescope} (\JWST) was designed to search for
faint galaxies at the highest redshifts. Its primary imager, the Near
InfraRed Camera (\JWST/NIRCam), will cover wavelengths in the range
$0.6 - 5 \mu m$ and is expected to observe some of the earliest stars
and galaxies \citep{beichman_science_2012}. These observations may
reveal the early stages of galaxy formation and provide an important
test of the $\Lambda$ cold dark matter ($\Lambda$CDM) model of the
universe, which predicts that galaxies are assembled hierarchically
starting from small, faint fragments that form at high redshift.

Theoretical predictions are vital for the interpretation of the upcoming observations. Cosmological hydrodynamical simulations and semi-analytic
modelling are the tools commonly employed for making such predictions
\citep[e.g.][]{tacchella_redshift-independent_2018,cowley_predictions_2018,yung_semi-analytic_2019,vogelsberger_high-redshift_2020}. For
example \citet{cowley_predictions_2018} and
\citet{yung_semi-analytic_2019} used semi-analytic modelling to
predict galaxy luminosity functions for \JWST/NIRCam passbands at various
redshifts. Similarly, \citet{vogelsberger_high-redshift_2020} used the
IllustrisTNG hydrodynamical simulations for the same purpose and
provided tailored predictions for two \JWST\ surveys: the \JWST\ Advanced Deep
Extragalactic Survey (JADES) and the Cosmic Evolution Early Release
Science Survey (CEERS). Hydrodynamical simulations like this have the
advantage that they can resolve the spatial distribution of gas in
galaxies allowing the effects of dust to be calculated in
post-processing. Several estimates already exist of the bright end of
the luminosity function \citep[e.g.][]{oesch_most_2014, finkelstein_evolution_2015, bouwens_uv_2015, bouwens_new_2021}; \JWST\ will extend these measurements to much fainter magnitudes.

As the best studied galaxy in the Universe, the Milky Way (MW) holds a
special place in studies of galaxy formation and evolution. Recent
advances, largely driven by data from the \textit{Gaia} satellite
\citep[][]{gaia_collaboration_gaia_2018}, have painted a much more
detailed picture of its assembly history than we had even a few years
ago. In particular, a major accretion event, in which a large dwarf
galaxy merged into the main progenitor was recently discovered, the
`Gaia-Enceladus' \citep{helmi_merger_2018} or `Gaia Sausage'
\citet{belokurov_co-formation_2018} (hereafter GES). Another large
accretion event that has been known for a long time is that of the
Large Magellanic Cloud (LMC), now known to be a very massive
satellite, with about 10 percent of the MWs mass
\citep[e.g][]{penarrubia_timing_2016,erkal_total_2019}. A massive accretion such as the LMC has been shown to be important when interpreting the satellite population of the MW and their orbital properties \citep{patel_orbits_2017,patel_orbits_2017-1}. There are several other suggested merger events present in the MW's history, these events tend to be either lower in stellar mass or at higher redshifts and are not very well characterised \citep[e.g.][]{kruijssen_kraken_2020, forbes_reverse_2020, naidu_evidence_2020,naidu_reconstructing_2021, horta_evidence_2021}.

The GES was discovered in {\em Gaia} chemo-dynamical data for the
inner Galactic halo by two groups\footnote{There is some debate as to
  whether or not these are the same event
  \citep[e.g.][]{elias_cosmological_2020,evans_early_2020}.}. This stellar component is thought to be the remnant of the merger of a relatively massive dwarf
galaxy ($\Mstar \sim 10^8-10^9 \Msun$) with the MW's progenitor about
$8-11$ Gyr ago which formed the majority of the galactic inner halo
and left a debris of stars on highly radial orbits
\citep[e.g.][]{fattahi_origin_2019, mackereth_origin_2019,
  amorisco_contributions_2017}. As shown by \citet{evans_how_2020}, if
the GES and LMC are the only massive ($\gtrsim 5\times10^8 \Msun$) accretion events, the MW's
accretion history would be unusually quiet for a galaxy of this mass
in the $\Lambda$CDM model. The presence of the LMC is also
exceptional: as first shown by \cite{benson_effects_2002} only
$\sim 10$ percent of MW analogues in $\Lambda$CDM simulations have
satellites as massive as the LMC \citep[see
also][]{boylan-kolchin_too_2011, busha_mass_2011, liu_how_2011,
  tollerud_large_2011}. 


In this work we analyse MW analogues identified in the EAGLE
cosmological hydrodynamics simulations
\citep{schaye_eagle_2015,crain_eagle_2015}. Our goal is to make
predictions for the properties of their progenitors that are, in
principle, accessible to the \JWST. We will consider the progenitors of average MW-like haloes (defined at $z=0$), as well as those of MW-analogues constrained by their accretion history. \citet{evans_how_2020} found that these analogues have lower mass at early times compared to average MW-like haloes, selected at $z=0$. We also investigate the properties of the progenitors of the LMC and GES analogues. 



This paper is organised as follows. The simulations and details of our
definitions of MW, LMC and GES analogues are discussed in
Section~\ref{sec: methods}. Our calculation of galaxy luminosities in
the \JWST/NIRCam passbands and the dust model we adopt are described in
Section~\ref{sec: mags}, where we also present predictions for
properties of the overall galaxy population, such as the luminosity
function at various redshifts.  Results for the MW, LMC and GES
progenitors are presented in Section~\ref{sec: mw results}.  Our paper
ends with a discussion of our main results and our conclusions in
Section~\ref{sec: disc and conc}.

\section{EAGLE simulations}
\label{sec: methods}

The EAGLE project consists of a set of cosmological hydrodynamical
simulations that follow the formation and evolution of galaxies in
large periodic cosmological volumes
\citep{schaye_eagle_2015,crain_eagle_2015}. The simulations were run
using a highly modified version of the smooth particle hydrodynamic
Tree-PM code {\sc p-gadget3}, which is based on the publicly available
{\sc gadget2} code, \citep{springel_cosmological_2005}. A full
description of the galaxy formation model is presented in
\citet{schaye_eagle_2015}. In short, it includes homogeneous UV-Xray
background radiation, metallicity dependant star formation and
cooling, stellar evolution and feedback, supermassive blackhole
accretion and AGN feedback. The EAGLE model has been shown to
reproduce many key features of the observed galaxy population, such as the stellar mass function at $z=0.1$ and realistic sizes down to
$\sim 10^8\Msun$, and produce galaxies with realistic mass profiles and rotation curves \citep[see][]{schaller_baryon_2015}. Also properties of MW-like galaxies in EAGLE have been shown to reproduce key features of our Galaxy \citep[e.g.][]{mackereth_origin_2019,thob_relationship_2019,evans_how_2020}.

The {\sc Friends-of-Friends} algorithm \citep{davis_evolution_1985}
was used, with a linking length of $0.2 \times$ the mean interparticle separation, to identify dark matter haloes. The {\sc subfind} algorithm \citep{springel_cosmological_2005} iteratively finds the substructure and
subhaloes within the {\sc Friends-of-Friends} groups. The adopted
cosmological parameters are based on the
\citet{planck_collaboration_planck_2014}; $\Omega_m=0.307$,
$\Omega_\lambda=0.693$, $\Omega_ {bar}=0.048$,
$H_0=67.77\,{\rm km\,s^{-1} Mpc^{-1}}$, $\sigma_8=0.8288$.

Unless otherwise stated, we use the fiducial EAGLE run which has a
periodic cubic volume of (100Mpc)$^3$ and was run with the `REFERENCE'
parameters \citep[REF-L0100N1504 in the nomenclature of][]{schaye_eagle_2015}. The initial mass for gas and matter particles are $9.6\times10^6~\Msun$ and $1.81\times10^6~\Msun$,
respectively. For convergence checks at the low mass, we use an EAGLE
run with 8$\times$ better mass resolution, but in a smaller volume,
(50Mpc)$^3$, which has been simulated with the `RECAL' parameters
(Recal-L0050N1504) run from the Exploring Neutral Gas in EAGLE
(ENGinE) simulations (Sykes et al. in prep)\footnote{this simulation
  has the same resolution as L025N0752 run in
  \citet{schaye_eagle_2015}. We do not use the latter due to the small
  size of the box and much fewer number of galaxies}. This simulation
has been run only up to $z=2$. For simplicity we refer to these runs
as `EAGLE-Ref' and `EAGLE-Recal', hereafter.

The stellar mass ($\Mstar$) of galaxies adopted in this work is calculated by summing the masses of bound star particles within 30kpc of the centre of galaxies. A 30kpc radius is appropriate for MW-mass galaxies at redshift $z=0$, since the majority of the stars are within this radius. At higher redshifts, where galaxies are smaller, this  boundary will include all of the particles in the galaxy. Unless mentioned otherwise, we include galaxies with stellar mass above $10^7 \Msun$, corresponding to $N\sim5$  and $44$ star particles in the EAGLE-Ref and EAGLE-Recal runs, respectively. The stellar masses of EAGLE-Ref galaxies have been shown to converge down to $N\sim5$ particles in \citet[][]{sawala_apostle_2016}.

Several element abundances, including Iron and Hydrogen are tracked
self-consistently in the simulations for gas and star particles. We
convert those mass fractions to [Fe/H] assuming a solar abundance of
$12+\log_{10}(N_{Fe}/N_H)=7.5$ from \citet{asplund_chemical_2009} to
assign magnitudes to each star particle (described in more detail in
section \ref{sec: mags}).

\subsection{Analogue definitions}\label{sec: mw def}
In this work we make use of many different galaxy ``groups'', and thus
provide clear definitions below. We define a MW-like galaxy as any
galaxy in EAGLE-Ref with halo mass in the range
$M_{200}=(0.7-2)\times 10^{12}\Msun$ \citep[see][and references
therein]{callingham_mass_2019}. `LMC-like' group includes satellites, located inside $R_{\rm200}$\footnote{The spherical radius with mean enclosed density
  200 times the critical density of the universe} of any MW-like galaxy at $z=0$, and have stellar masses in the range
$\Mstar =(1-4)\times 10^9\Msun$. `GES-like' galaxies are any galaxies which have a
stellar mass of $\Mstar = (0.5-1)\times 10^9 \Msun$ when they merge
with a MW-like galaxy between redshift $z=1$ and 2 (8-10 Gyr ago). Note that we do not place any constraints on having a Local Group environment, which could affect the formation epoch of our haloes \citep[][]{santistevan_formation_2020}. 

Our `MW analogues' are MW-like galaxies with additional constraints on their accretion history, following \citet{evans_how_2020}:
\begin{itemize}
    \item one LMC-like satellite present at $z=0$ with no other more massive satellites
    \item one GES-like merger event with no more massive mergers within the same time-frame
    \item finally, we require that these systems have a `merger free
      zone' when there is an absence of massive mergers ($\Mstar>0.5
      \times 10^9\Msun$) between redshifts $z=0$ and $z=2$. 
\end{itemize}

The definition of LMC satellites and GES mergers are deliberately broad in the hope of having better statistics. Table \ref{table: basic properties} gives the number of galaxies in each of the groups used throughout this paper and their median stellar masses at redshifts $z=0$ and $z=2$. More specific properties of the MW analogue systems are presented in \citet[][]{evans_how_2020}.

\begin{table}
\centering
\caption{The number of galaxies in each of the galaxy groups studied in this work and their median $z=0$ and $z=2$ stellar masses.}
\renewcommand\arraystretch{1.5}
\begin{tabular}{@{}llll@{}}
\toprule
Group        & Number & Median M$_{*,z=0}$ & Median M$_{*,z=2}$ \\
       &  &  [$\times10^9\Msun$] & [$\times10^8\Msun$] \\\midrule
MW-like      &  1078   &        20.3            &           23.8     \\
LMC-like     &  169    &       1.89             &        1.96          \\
GES-like     &  234    &        --            &       6.40         \\ \midrule
MW-analogue  &  7      &       14.4             &      10.4              \\
LMC-analogue &  7      &         1.29           &     2.10               \\
GES-analogue &  7      &         --           &         4.79           \\ \bottomrule
\end{tabular}
\label{table: basic properties}
\end{table}

\section{Galaxy luminosities and colours}
\label{sec: mags}
In this section we describe how we calculate the dust-free magnitudes
of the simulated galaxies for \JWST/NIRCam passbands, as well as the
absolute rest-frame UV. We also describe the model adopted throughout
this work to account for dust attenuation.

\subsection{Dust-free magnitudes}

We use the initial mass function (IMF), age and metallicity of
simulated star particles, combined with publicly available stellar
libraries to retrieve their spectral energy distributions (SEDs). We use the Flexible Stellar
Population Synthesis (FSPS) code \citep{conroy_propagation_2009,
  conroy_propagation_2010} with the MESA Isochrones \& Stellar Tracks
\citep[MIST,][]{paxton_modules_2011, paxton_modules_2013,
  paxton_modules_2015, choi_mesa_2016, dotter_mesa_2016} and MILES
stellar library \citep{sanchez-blazquez_medium-resolution_2006}. The
IMF adopted in the simulations is Chabrier
\citep{chabrier_galactic_2003} with an initial mass range of $0.1-100 \Msun$. The
stellar isochrones cover the following range of age and metallicity:
$-2.5 \leq {\rm [Fe/H]} \leq 0.5$ with 12 intervals, and
$5 \leq {\rm log(age/yrs)} \leq 10.3$ with 107 equally spaced points
(these intervals were predetermined by FSPS). We identify the
isochrone with the nearest metallicity and age to the stellar
particles. If any of the star particles lie outside the
age-metallicity grid of isochrones, they are also assigned to the
nearest isochrone\footnote{We have checked that only a small fraction (3\%) of star particles fall outside this grid at $z=4$. Hence they would have negligible effect on the overall luminosity of each galaxy.}.

The magnitudes in various passbands are retrieved by applying the response
of each filter to the SED, which is done automatically by FSPS. For
galaxies at higher redshifts ($z>0$), the SED is redshifted before
applying the filter. The total magnitude of each galaxy is
calculated by adding the flux of all bound star particles within $r<30$kpc.

\subsection{Dust model}

We compute the dust attenuation for each star particle in the
simulated galaxies using a semi-empirical approach, following a modified version of `model B' in
\citet{vogelsberger_high-redshift_2020}. The modification accounts for
the fact that the gas component in the EAGLE simulation is represented by SPH
particles, rather than Arepo's Voronoi mesh cells in IllustrisTNG. We
smooth the gas particles over a cubic grid, as detailed below.  There
are two different components to the dust model: resolved dust from the
ISM, and unresolved dust from stellar birth clouds

\begin{figure*}
    \centering
    \includegraphics[width=0.84 \textwidth]{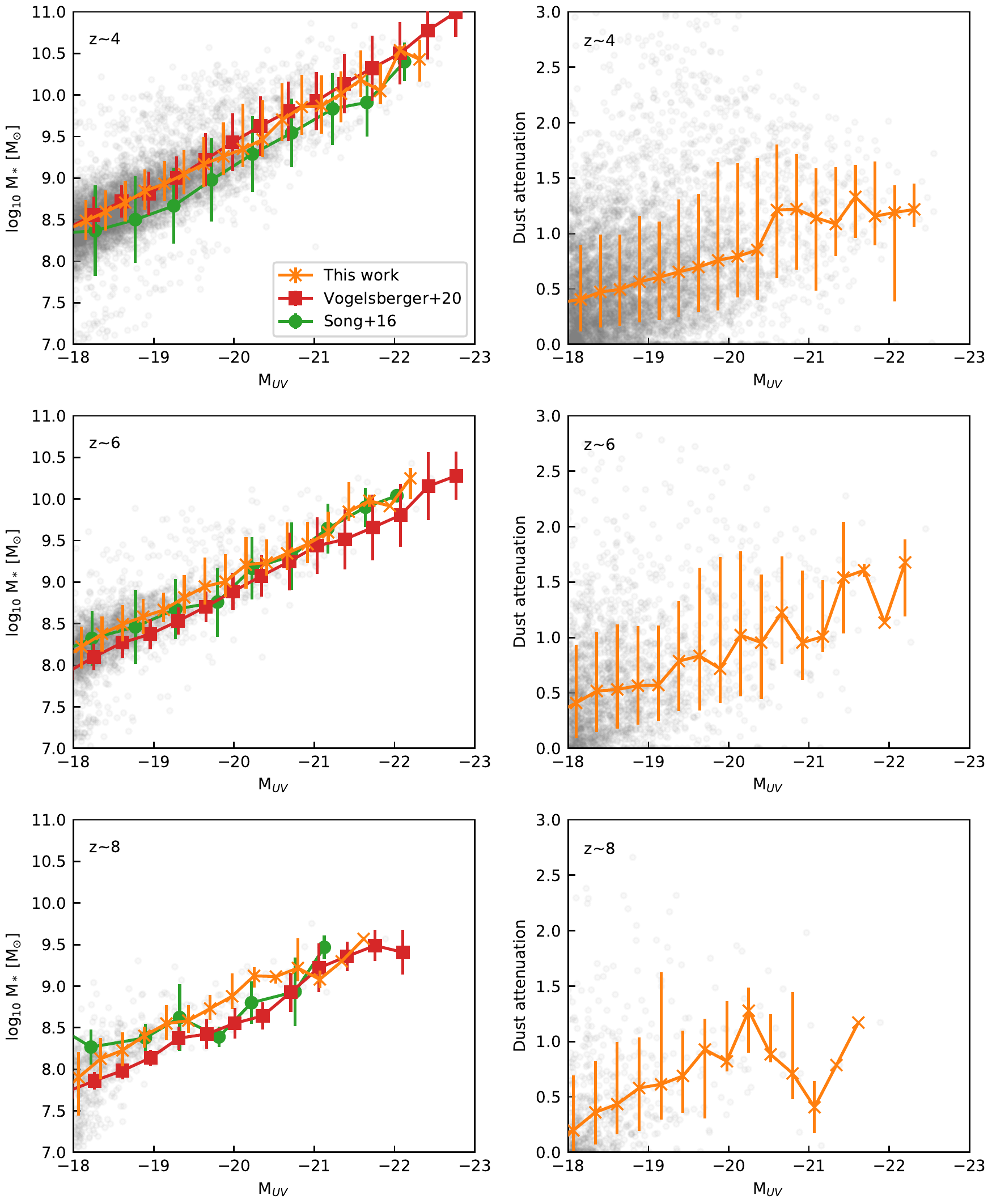}
    \caption{\textit{Left column}: Stellar mass versus dust-corrected rest-frame UV magnitude at redshifts $z=4, 6, 8$, with grey scatter points
      corresponding to individual galaxies from EAGLE-Ref and the
      orange line showing median and [$16^{th}-84^{th}$] percentile at
      a fixed $M_{\rm UV}$. The green connected circles and red
      connected squares show the results from the full radiative dust
      model in \citet[][]{vogelsberger_high-redshift_2020}, and
      observations from \citet[][]{song_evolution_2016},
      respectively. \textit{Right column}: Dust attenuation as a
      function of magnitude at redshifts $z=4, 6, 8$, for galaxies in
      EAGLE-Ref. Our analytic dust model produces comparable results
      to SKIRT and observations.}
    \label{fig: dust compare}
\end{figure*}

\subsubsection*{Resolved dust}

The resolved dust attenuation is caused by cold ($<10^4$K) or
star-forming gas in the interstellar medium (ISM) along the
line-of-sight. Unless otherwise stated, the line of sight direction
for each galaxy is random. We use the smoothing lengths of gas
particles to smooth the density using the original EAGLE kernel, over
a cubic grid with 1 kpc spacing. We then carry out the following
calculations \citep[for more detailed information see][section
3.2.2]{vogelsberger_high-redshift_2020} to obtain the attenuation for
each star particle,

\begin{equation}
    \tau_V^{\text{res}} = \tau_{\text{dust}}(z) \biggl(\frac{Z_g}{Z_{\odot}}\biggl)^{\gamma}\biggl(\frac{N_H}{N_{H,0}}\biggl)
    \label{eqn: tauV}
\end{equation}
where $N_{H}$ is the hydrogen column density along the line of sight `in front' of the star particle, $\tau_{\text{dust}}(z)$ is the redshift dependent scale factor for the optical depth which scales as the average dust-to-metal ratio, $\gamma=1$ and $N_{H,0} = 2.1 \times 10^{21}$cm. The V-band optical depth, $\tau_V$, values are then converted into the V-band dust attenuation using the following relation,

\begin{equation}
    A_V^{\text{res}} = -2.5 \log \biggl(\frac{1-e^{-\tau_V^{\text{res}}}}{\tau_V^{\text{res}}}\biggl).
    \label{eqn: tauV to AV}
\end{equation}

Since the optical depth and dust attenuation are both specific to the V-band, they need to be converted to the optical depth and dust attenuation for the passbands we are interested in (\JWST/NIRCam and absolute rest-frame UV). To convert from V-band attenuation to attenuation for a given wavelength, $\lambda$, we adopt the \citet{calzetti_dust_2000} relation \citep[modified by][to include the UV bump]{kriek_dust_2013} for local starburst galaxies such that:

\begin{equation}
    A^{\text{res}} (\lambda) = \frac{A_V^{\text{res}}}{4.05} [k^{\prime}(\lambda) + D(\lambda)] \biggl(\frac{\lambda}{\lambda_V} \biggl)^{\delta}
    \label{eqn: calzetti attenuation relation}
\end{equation}
where $k^{\prime}(\lambda)$ is the normalised attenuation curve for $A_V$:

\begin{equation}
    k^{\prime}(\lambda) = 4.05 + 2.659 \left\{\begin{matrix}
\biggl(-1.875 +\frac{1.040}{\lambda}\biggl), \\ 
\text{for } 0.63\mu m < \lambda < 2.20\mu m\\ \\
\biggl(-2.156 +\frac{1.509}{\lambda} -\frac{0.198}{\lambda^2} +\frac{0.011}{\lambda^3}\biggl), \\ 
\text{for } 0.12\mu m < \lambda < 0.63\mu m
 \end{matrix}\right.
\end{equation}
and $D(\lambda)$ parameterises the UV bump which is given by:
\begin{equation}
    D(\lambda) = \frac{E_b(\lambda\Delta\lambda)^2}{(\lambda^2-\lambda_0^2)^2+(\lambda\Delta\lambda)^2}
\end{equation}
where $\lambda_0 = 217.5$ nm and $\Delta\lambda=35$ nm are the central wavelength and full width half maximum (FWHM) of the UV bump respectively \citep{seaton_interstellar_1979,noll_analysis_2009}. The shape of the attenuation curve is purely characterised by $\delta$ as shown by the relation between $E_b$ and $\delta$ found by \citet{kriek_dust_2013}:

\begin{equation}
    E_b = (0.85 \pm 0.09) - (1.9\pm 0.4)\delta
\end{equation}
we assume $\delta=0$ in order to apply no correction to the attenuation curve other than the addition of the UV bump as in \citet{vogelsberger_high-redshift_2020}. The overall correction for the magnitude for the resolved dust component, in any given filter, is therefore:
\begin{equation}
    M^{\text{dust}} = M^{\text{dust-free}} + A^{\text{res}}(\lambda).
\end{equation}

\subsubsection*{Unresolved dust}
The unresolved dust component of the model accounts for the stellar
birth clouds around young stars which are not resolved in EAGLE. We
include this component by assuming that all star particles in a given
galaxy will have the same dust attenuation from their birth
clouds\footnote{Since we are only interested in the dust attenuation for galaxies as a whole, dust attenuation values for individual star particles are not as important.}. The birth cloud V-band optical depth is given by: 

\begin{equation}
\tau_V^{\text{unres}}=\left\{\begin{matrix}
2 \langle \tau_V^{\text{res}} \rangle, & \text{for } t' \leq t_{\text{disp}}\\ 
0, & \text{for } t' > t_{\text{disp}}
 \end{matrix}\right.
\end{equation}
where $\langle \tau_V^{\text{res}} \rangle$ is the average V-band
optical depth of the whole galaxy (computed using Eqn. \ref{eqn:
  tauV}) and t$_{\text{disp}} =$ 10 Myr is the dispersion time for the
stellar birth cloud. Hence, if a star particle is younger than the
dispersal time of the stellar birth cloud then all star particles
satisfying this criteria will have the same additional optical depth
value. Again, the optical depth needs to be converted to the
attenuation, here we assume a simple uniform dust screen such that the
solution for the radiative transfer equation takes the following form: 
\begin{equation}
    \begin{split}
A^{\text{unres}}_V & = -2.5\log(e^{-\tau_V^{\text{unres}}}) \\
 & = 1.086 \text{ } \tau_V^{\text{unres}}.
\end{split}
\end{equation}
The dust attenuation at other wavelengths is estimated using a simple
power law relation from \citet{charlot_simple_2000} for unresolved
dust: 
\begin{equation}
    A^{\text{unres}}(\lambda) = A^{\text{unres}}_V \biggl(\frac{\lambda}{\lambda_V}\biggl)^{-0.7}.
\end{equation}
Combining the resolved and unresolved dust then gives the total
magnitude correction, in any filter, such that: 
\begin{equation}
    M^{\text{dust}}=M^{\text{dust-free}} + A^{\text{res}}(\lambda) + A^{\text{unres}}(\lambda).
\end{equation}

We show the high redshift M$_{UV}$-stellar mass relation of simulated
galaxies from the EAGLE-Ref run in the left column of Fig.~\ref{fig: dust compare} after
applying dust attenuation, and compare them with the results of
\citet[][]{vogelsberger_high-redshift_2020} from IllustrisTNG, as well
as observations from \citet{song_evolution_2016}. \citet{song_evolution_2016} analysed data from the \textit{Hubble Space Telescope} which included $\sim 7000$ galaxies selected using photometric redshifts in the range $z=3.5-8.5$; further details may be found in \citet[][and references therein]{song_evolution_2016}. Grey points are
individual galaxies and the orange curve with error bars shows the
median stellar mass and the $[16^{th}-84^{th}]$ percentiles at fixed
magnitude. We only show results for redshifts $z=$4, 6 and 8, for
which data from \citet{song_evolution_2016} are available. Our results
are in excellent agreement with those of \citet[][model
`C']{vogelsberger_high-redshift_2020} which is a more comprehensive
and computationally expensive dust model using the radiative transfer
method SKIRT \citep[][]{baes_efficient_2011, camps_using_2013, saftly_hierarchical_2014, camps_skirt_2015}. Our dust model uses additional information from the particles in the simulation (unlike simple empirical models) and shows quantitatively similar results to the full radiative dust model (SKIRT). This is very reassuring that they show such excellent agreement. EAGLE galaxies are also consistent with observational data within the scatter. The right column of Fig.~\ref{fig: dust compare} shows the dust attenuation as a function of M$_{UV}$ for redshifts $z=$4, 6 and~8, indicating that our dust attenuation increases by approximately $\sim 1$ mag as M$_{UV}$ magnitude changes from -18 to -22, this result is consistent with \citet[][]{yung_semi-analytic_2019}. This is due to brighter (more massive) galaxies having a larger amount of (cold) gas. We also compared our dust-corrected magnitudes with those of \citet[][]{trayford_colours_2015} who calculated the dust-free and dust-corrected SDSS apparent magnitudes for EAGLE-Ref galaxies using SKIRT at $z=0.1$. Our results are consistent with theirs in the mass range $\Mstar>1\times10^9\Msun$.


\subsection{Luminosity functions}  
\label{sec:lumfunc}

\begin{figure}
    \centering
    \includegraphics{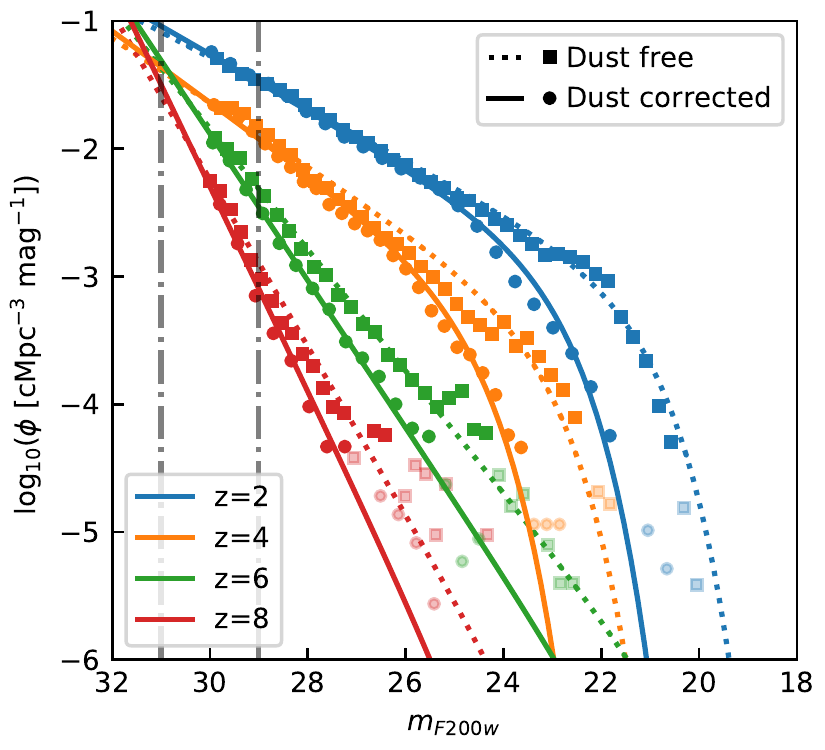}
    \caption{Luminosity functions of galaxies in the EAGLE-Ref
      simulation at different redshifts in the \JWST/NIRCam F200W
      passband. The dust-free and dust-corrected luminosity functions are
      shown as square and circular points, respectively, with their
      corresponding Schechter fits as dotted and solid
      lines. Different redshifts are highlighted with different
      colours, as shown in the legend. Open faded symbols at the brighter end highlight bins with fewer than 10 galaxies per bin. The vertical dashed-dotted
      lines at limiting magnitudes of $m_{\rm lim} = 29$ and $31$ mag
      correspond to the faintest magnitudes that are observable with
      exposure times of $T_{\rm exp} = 10^4$s and $10^5$s,
      respectively.}
    \label{fig: dust corrections}
\end{figure}

\begin{figure}
    \centering
    \includegraphics[width=0.95\columnwidth]{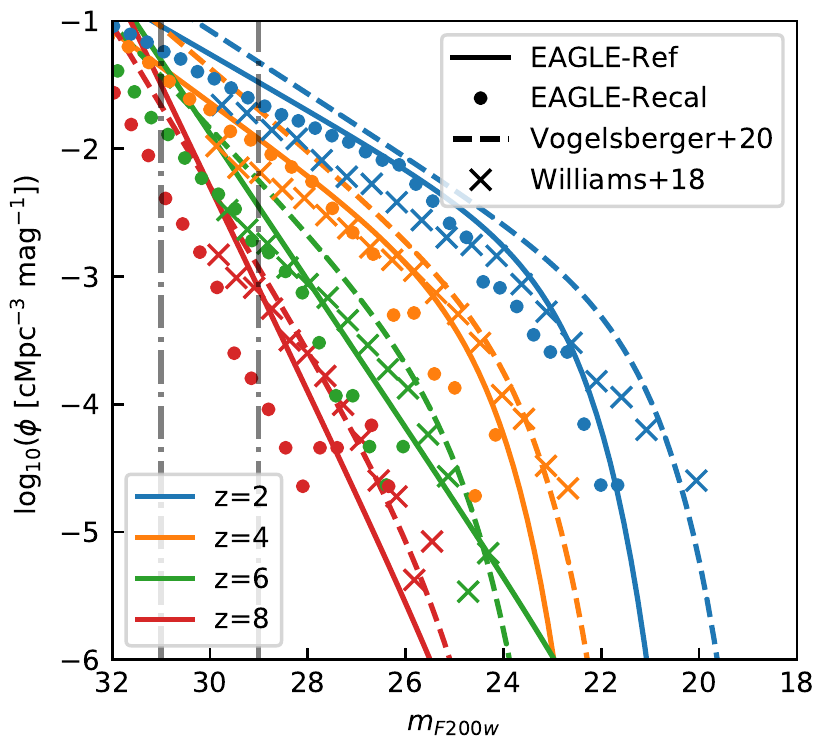}
    \caption{Dust-corrected luminosity functions for the \JWST/NIRCam F200W passband for redshifts $z=2,\ 4,\ 6, \ 8$. Solid lines correspond to the Schechter fits for EAGLE-Ref luminosity functions (same as Fig.~\ref{fig: dust corrections}), whereas circles of the same colour show the higher resolution EAGLE-Recal luminosity functions. For comparison, the dashed lines represent the luminosity functions from the Illustris-TNG simulations, computed using radiative transfer dust model and presented in \citet{vogelsberger_high-redshift_2020}. Crosses show the luminosity functions derived from the JAGUAR mock catalogue for \JWST\ \citep[][]{williams_jwst_2018}. The vertical dashed-dotted lines are the same as in Fig.~\ref{fig: dust corrections}.}  
    \label{fig: luminosity function}
\end{figure}

Luminosity functions give the comoving number density of galaxies at a
given luminosity; they are typically represented by a Schechter
function \citep{schechter_analytic_1976} with the following form in
magnitude space,

\begin{equation}
    \phi(m) = \frac{0.4\ln(10)\phi^*}{10^{0.4(m-m^*)(\alpha^*+1)}} \exp(-10^{-0.4(m-m^*))}),
    \label{eqn: schechter}
\end{equation}
where $\phi^*$ is the normalisation, M$^*$ is the transition
magnitude, and $\alpha^*$ is the faint-end slope parameter. 

Fig.~\ref{fig: dust corrections} shows the comoving luminosity
function of the simulated galaxies before and after dust correction,
and the corresponding Schechter fits in the \JWST\ F200W passband. The best-fit parameters for the Schechter function were calculated using a
$\chi^2$ method for magnitudes brighter than 30, with Poisson
uncertainties (the best-fit parameters for the \JWST\ F200W passband are presented in Table \ref{table: schechter parameters}).   

\begin{table}
\centering
\caption{The best-fit Schechter parameters for the \JWST\ F200W passband at redshifts $z=2,4,6,8$.}
\renewcommand\arraystretch{1.5}
\begin{tabular}{@{}llll@{}} 
\toprule
Redshift        & $\phi^*$ & m$^*$ & $\alpha^*$ \\ 
&[cMpc$^{-3}$ mag$^{-1}$]&[mag]&\\\midrule
2      &  0.0018 & 23.09 & -1.55   \\
4     &  0.00091    &      24.83              &       -1.70           \\
6     &  $1\times10^{-7}$&        21.03          &     -2.44         \\ 
8  &   $1\times10^{-7}$      &       24.06             &       -2.99             \\\bottomrule
\end{tabular}
\label{table: schechter parameters}
\end{table}

Fig. \ref{fig: dust corrections} shows that including dust affects the bright end of the luminosity function more than the faint end. This is expected according to the right column of Fig.~\ref{fig: dust compare}. Moreover, Fig.~\ref{fig: dust corrections} shows that the dust has a larger impact at lower redshifts. This is expected since the average metallicity of galaxies is higher at lower redshifts due to past star formation.

Fig.~\ref{fig: luminosity function} shows the dust-corrected
luminosity function of the EAGLE galaxies at two resolution levels,
alongside the results of Illustris-TNG
\citep{vogelsberger_high-redshift_2020}.  The solid line corresponds
to a Schechter fit to galaxies in the EAGLE-Ref simulation (repeated from Fig.~\ref{fig: dust corrections}); points show the higher
resolution results from EAGLE-Recal. Small differences between the luminosity
functions, EAGLE-Ref and EAGLE-Recal, are expected, as the two models
have slightly different parameters \citep[see,][for comparison of
stellar mass functions at $z=0$]{schaye_eagle_2015}. Our results,
however, indicate that the low mass end slope of our Schechter fit is
not significantly affected by the lower resolution of the EAGLE-Ref
run for redshifts $z<4$. At higher redshifts, $z=6,8$, the differences between EAGLE-Ref and EAGLE-Recal become larger but these differences are still consistent within Poisson uncertainties (see e.g. Fig.~\ref{fig: nexp vs redshift}).
The increasing difference between EAGLE-Ref and EAGLE-Recal is due to the slight differences in the parameters of subgrid models \citep[see][for more information]{schaye_eagle_2015}.

The dashed lines in Fig.~\ref{fig: luminosity function} show the
outcome of model C, a full radiative transfer dust approximation using
SKIRT presented by \citet{vogelsberger_high-redshift_2020}. Despite
the good agreement of the $\Mstar-M_{\rm UV}$ relation between our
results and those of \citet{vogelsberger_high-redshift_2020},
especially at $z<5$, shown in Fig.~\ref{fig: dust compare}, there are
notable differences in the luminosity functions. This implies that the
difference is mainly coming from the differences in the stellar mass
functions, or equivalently stellar mass-halo mass relations, between
the two sets of simulations. The largest difference is seen at the
brighter end and at lower redshift, so it is likely due to the
differences in AGN models and feedback. We note that
\citet{vogelsberger_high-redshift_2020} used a combination of
IllustrisTNG volumes; the largest one (TNG-300) is $\approx 30 \times$
larger than the EAGLE-Ref volume and therefore better samples the
bright end of the luminosity function. We show bins with fewer than 10
galaxies as faint points in Fig.~\ref{fig: dust corrections}. The crosses in Fig.~\ref{fig: luminosity function} show the luminosity functions derived from JADES Extragalactic Ultra-deep Artificial Realization \citep[JAGUAR;][]{williams_jwst_2018}. The foundations of the JAGUAR mock catalogue were constructed using observations from \citet[][]{tomczak_galaxy_2014} and extrapolated to match the UV luminosity functions in \citet[][]{oesch_probing_2013, bouwens_uv_2015, bouwens_bright_2016, calvi_bright_2016, stefanon_hst_2017, oesch_dearth_2018}. The luminosity functions from JAGUAR agree well with our results, however, the luminosity functions are flatter throughout. Thus, the EAGLE simulations might underestimate the number of bright galaxies and overestimate the number of faint galaxies which could be observed with \JWST/NIRCam. The flattening of the faint end slope in the \citet[][]{williams_jwst_2018} data is more pronounced at higher redshifts. This could be a result of the increasing difference in $\alpha^*$ values in the Schechter functions. Our $\alpha^*$ is consistent with \citet[][]{bouwens_uv_2015} at $z=4$ who estimate the slope for the UV luminosity function to be $-1.67\pm0.05$; however, at redshift $z\sim8$ their slope is at least $\Delta\alpha^*\sim0.9$ flatter.

\subsection{Number of galaxies in \JWST/NIRCam field-of-view}

\begin{figure}
    \centering
    \includegraphics{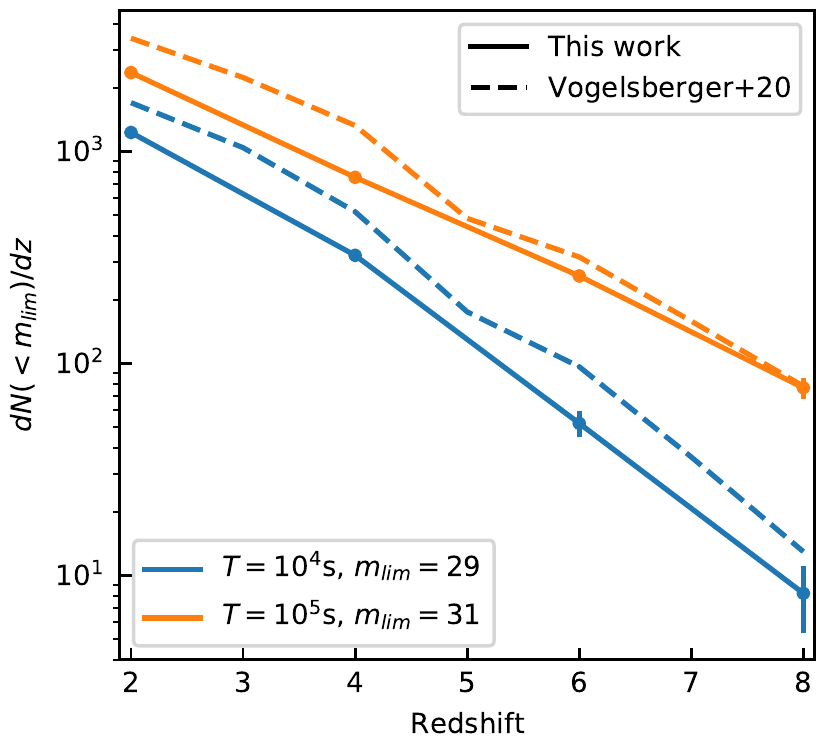}

    \caption{The expected number of galaxies as a function of redshift,
      in a \JWST/NIRCam FoV ($2.2\times 2.2$ arcmin) that are above the
      detection limit with exposure times of $T=10^4$s and $T=10^5$s,
      and $SNR =10$ and 5 respectively. These exposure times translate
      to limiting magnitude of $m_{\rm lim} = 29$ and $31$,
      respectively. The solid and dashed lines correspond to our
      EAGLE-Ref results, and those of
      \citet{vogelsberger_high-redshift_2020}, respectively. Error
      bars show the Poisson error on each value.} 
    \label{fig: nexp vs redshift}
\end{figure}

Our predictions for the luminosity function of galaxies can be used to
estimate the number of galaxies observable within a \JWST/NIRCam
field-of-view (FoV). We need to integrate the Schechter fits, as in
Eqn.~\ref{eqn: cumulative schechter}, above the observable magnitude
limit:

\begin{equation}\label{eqn: cumulative schechter}
\begin{split}
\phi_{\text{cum}}(<m_{\text{lim}}) & = \int_{L_{\text{lim}}}^\infty \phi(L)dL \\
 & = \phi^* \Gamma_{\text{inc}}(\alpha^* +1, 10^{-0.4(m_{\text{lim}}-m_*)}),
\end{split}
\end{equation}
where $\alpha^*, \phi^*$ and $m^*$ are the parameters of the Schechter
function, and $\Gamma_{\text{inc}}(a,z) = \int_z^{\infty} t^{a-1}
e^{-t} dt$ is the upper incomplete gamma function; $m_{\rm lim}$
represents the magnitude limit which depends on the exposure time and
signal-to-noise ratio ($SNR$). The limiting magnitudes used correspond to
exposure times of $T=10^4$s and $T=10^5$s with $SNR =10$ and 5
respectively; these result in $m_{\rm lim} = 29,31$. $m_{\rm
    lim}\sim$29 corresponds to the expected limiting magnitude for the
  JADES-M survey.

Finally, the following relation can be used to compute the expected
number of galaxies per unit redshift in the \JWST/NIRCam FoV: 

\begin{equation}
    \frac{\text{d}N_{\text{exp}}}{\text{d}z} = \phi_{\text{cum}}(<m_{\text{lim}}) \frac{\text{d}V_{\text{com}}}{\text{d}\Omega \text{d}z}(z) \Delta\Omega,
\end{equation}
where $\text{d}V_{\text{com}}/\text{d}\Omega \text{d}z$ is the differential comoving volume element described in Eqn. \ref{eqn: comoving volume} and $\Delta \Omega$ is the solid angle produced by the \JWST/NIRCam FoV ($2.2 \times 2.2$ arcmin).
\begin{equation}
    \frac{\text{d}V_{\text{com}}}{\text{d}\Omega \text{d}z}(z) = \frac{c(1+z)^2d_A(z)^2}{H_0E(z)}
    \label{eqn: comoving volume}
\end{equation}
where $d_A$ is the angular diameter distance and $H(z)=H_0E(z)$ is the
Hubble parameter at redshift $z$.  

Fig.~\ref{fig: nexp vs redshift} shows our predictions for the
observable number of galaxies per unit redshift in the \JWST/NIRCam FoV for
magnitude limits of $m_{\rm lim} = 29$ and 31 (corresponding to the
detection limits for exposure times of $T=10^4$s and $T=10^5$s, and
$SNR =10$ and 5 respectively). The error bars represent the Poisson
error on each value.

Fig. \ref{fig: nexp vs redshift} indicates that our expected number of galaxies is lower than those predicted in \citet{vogelsberger_high-redshift_2020}, by roughly $N=500 (1000)$ at $z=2$ for the $T=10^4$s ($T=10^5$s) exposure time. This is due to the systematically higher offset in the luminosity function of \citet{vogelsberger_high-redshift_2020} compared to EAGLE at all magnitudes at $z=2$, as seen in Fig.~\ref{fig: luminosity function}. The same statement is true at redshift $z=4$; however, this differs for the luminosity functions at redshifts $z=6-8$ primarily between magnitudes 29 and 31 (vertical dashed dotted lines), thus only affecting our expected number of galaxies for an exposure time of $T=10^5$s (shown in orange; corresponding to limiting magnitude of 31). Our expected number of galaxies for $T=10^5$s becomes much closer to the predictions of \citet[][]{vogelsberger_high-redshift_2020} at high redshift which are only lower by $N\sim60 (1)$ at $z=6$ ($z=8$). We note that these differences are mainly driven by the faint end since the number of galaxies is dominated by galaxies in this regime. We also found that our predicted numbers of galaxies are consistent with \citet[][]{cowley_predictions_2018}, who used semi-analytic modelling techniques.

\section{Progenitors of MW, LMC and GES}\label{sec: mw results}

\begin{figure*}
    \centering
    \includegraphics[width=\textwidth]{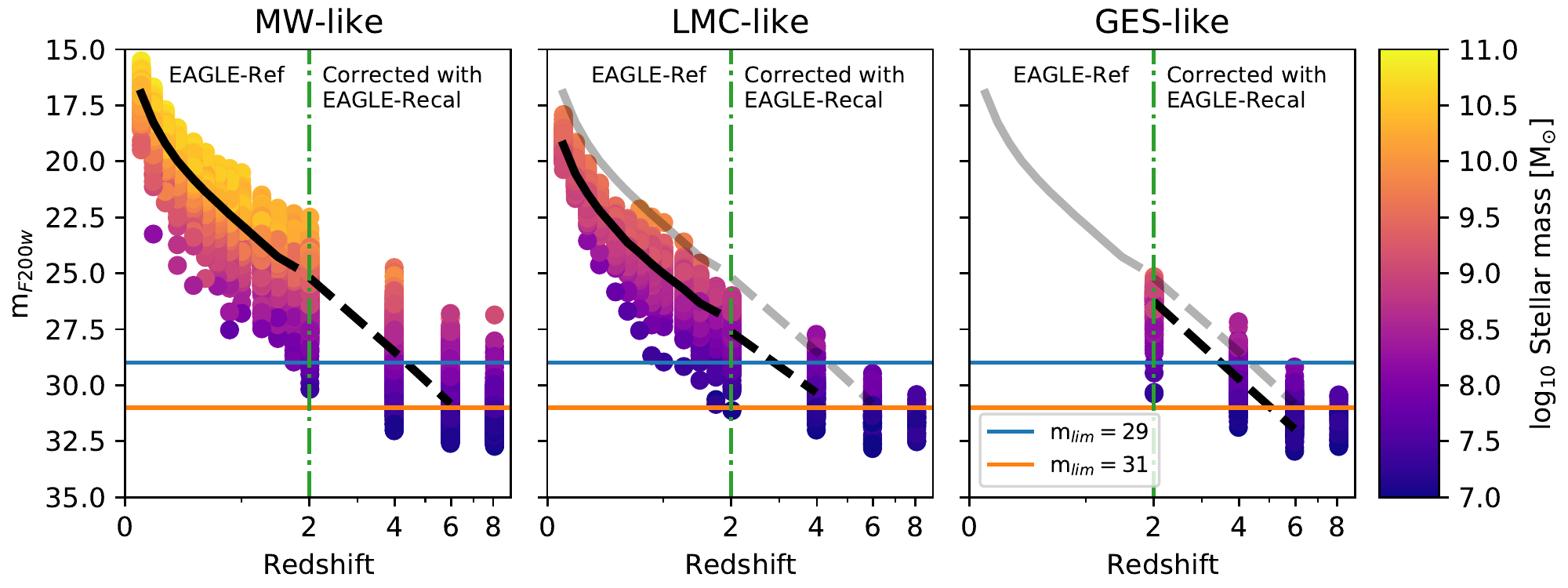}
    \caption{The apparent magnitude in the F200W passband as a function of
      redshift for progenitors of MW-like galaxies (\textit{left}),
      the LMC-like satellites (\textit{middle}) and GES-like galaxies
      (\textit{right}) selected from the EAGLE-ref simulations. The 
      points in each panel are coloured by the stellar mass, in logarithmic scale, of the
      galaxies as shown in the colour bar.  Points to the left of the
      vertical dash-dotted line ($z=2$) are magnitudes computed
      directly from EAGLE-Ref outputs, whereas magnitudes to the right
      of the line have been corrected using the $\Mstar-m_{\rm F200W}$ relation of the higher resolution EAGLE-Recal simulation (see Appendix
      \ref{sec: app mag correct} for details). The solid black lines in each panel
      show the median apparent magnitude at each redshift up until the
      boundary at $z= 2$, beyond which it turns into dashed indicating
      the transition to corrected magnitudes. The median line for the
      MW-like galaxies is repeated, as grey, in the middle and right
      panel for reference.  The two horizontal lines in blue and
      orange show the magnitude limits for exposure times of 10$^4$s
      and 10$^5$s respectively with a $SNR=10$ and 5, respectively.}
    \label{fig: mag vs red}
\end{figure*}

In this section, we focus on progenitors of MW analogues
that could be observed by \JWST. All the magnitudes and colours shown in this section
include dust attenuation. Our definition of MW-, LMC- and GES-like
galaxies, as well as MW analogues, are summarised in section \ref{sec:
  mw def}. 

Fig.~\ref{fig: mag vs red} shows the evolution of dust-corrected F200W
apparent magnitude as a function of redshift for the MW-like, LMC-like
and GES-like galaxies. Points are coloured according to their
stellar mass as shown in the colour bar, and the lines show the median
magnitudes at any given redshift. At lower redshifts $z<2$, we show
the magnitudes of progenitors, calculated directly from the EAGLE-Ref
run. At $z>2$, where stellar masses become smaller and resolution
effects become important, we correct the magnitudes statistically
using the stellar mass of the progenitors and the higher resolution
EAGLE-Recal run. Details can be found in Appendix \ref{sec: app mag
  correct}. We apply the correction only at $z>2$ and for progenitors
with $M_{*}< 10^{8}\Msun$, which is where our calculated magnitudes
show a large scatter at fixed stellar mass in the EAGLE-Ref run, due
to the limited resolution of the simulation. The median lines turn
from solid to dashed at $z>2$ when magnitudes have been corrected, and
the median line for the MW-like sample has been repeated in grey in
the other two panels for reference.

As expected, the progenitors are typically fainter at earlier times, 
albeit with significant scatter, which increases towards higher
redshift. This is particularly true for MW-like galaxies. For
example, the median magnitude and the interquartile range for MW-like
progenitors are $m_{\rm F200W}=20.8\pm0.58$ at $z\sim0.5$ and they
change to $25.2\pm0.84$ at $z\sim2$.

At redshifts higher than $z\sim3$, the fainter end of the magnitudes 
approach a constant value of $m_{\rm F200W}\sim 33$ mag. This is not
physical, and is due to the low mass progenitors not being identified
by the halo-finder at early times. In these circumstances, we show the
median assuming unidentified progenitors are all fainter than 
identified ones. We stop showing the median if more than 50 percent
of the progenitors in the sample are unidentified.

The two horizontal lines shown in Fig.~\ref{fig: mag vs red} indicate
the same detection limit of \JWST/NIRCam used in the previous section:
exposure times of $T=$10$^4$s and 10$^5$s, are shown as blue and
orange, respectively. The median of MW-like progenitors is easily
above the detection thresholds at $z<4$. However, the large scatter
causes the fainter progenitors to become undetectable from $z\sim2$.
LMC-like progenitors are on average fainter than the MW-like sample by
only $\sim 2$ mag at most redshifts, and the two samples overlap
significantly. The LMC-like sample is detectable on average to
$z\sim 2.8$ for $T=10^4$s with almost none detectable beyond
$z>4$. The redshifts when the median magnitudes reach detection
thresholds are summarised in Table \ref{table: mw max redshift}, for
various \JWST/NIRCam passbands. The maximum redshifts observable for the three
galaxy samples are all in the F356W passband, $ \sim 6, \sim 4, 5.3$
respectively. The passband with the lowest maximum redshift for the three
types of galaxies is the F070W passband. F356W is likely to be the most sensitive passband because it has the best transparency whereas F070W is likely to be the worst because of the lower flux at the blue-end of the spectrum, as well as a lower transparency.

GES-like galaxies, by definition, merge with their host MW-like
galaxy in the redshift range $z=1-2$, and therefore no data are shown
at $z<2$ for their progenitors in the right hand panel of
Fig.~\ref{fig: mag vs red}. Interestingly, the GES-like progenitor
sample is only slightly fainter than the MW-like progenitors ($\sim 1.1$
mag on average), and they are brighter than LMC progenitors. These
results are shown in more detail for $z=2$ in Fig.~\ref{fig:
  histograms mass mag}.

\begin{table}
\centering
\caption{The redshifts above which average progenitors of
  MW-like, LMC-like \& GES-like galaxies fall below the magnitude
  detection limit. Here we assume exposure times of $10^4$s and
  $10^5$s, with $SNR=10$ and 5 respectively for each of the
  \JWST/NIRCam photometric passbands.} 
\renewcommand\arraystretch{1.5}
\begin{tabular}{@{} c c c c c c c c @{}}
\toprule
 & &\multicolumn{2}{c}{MW mass}&\multicolumn{2}{c}{LMC mass}&\multicolumn{2}{c}{GES mass}\\ 
 & &$10^4$s         &    $10^5$s    &$10^4$s         &    $10^5$s&$10^4$s         &    $10^5$s    \\ \midrule
F070W && 4.1 & 4.9  & 2.4 & 4.0 & 3.2 & 4.4\\
F090W && 4.4 & 6.0  & 2.5 & 4.0 & 3.3 & 4.8\\ 
F115W && 4.3 & 6.0  & 2.7 & 4.0 & 3.3 & 4.8\\ 
F150W && 4.3 & 6.0  & 2.9 & 4.0 & 3.3 & 4.9\\ 
F200W && 4.5 & 6.0  & 3.0 & $\sim 4$ & 3.5 & 5.0\\ 
F277W && 4.5 & 6.0  & 3.0 & $\sim 4$ & 3.5 & 5.0\\ 
F356W && 4.9 & $\sim 6$ & 3.4 & $\sim 4$ & 4.0 & 5.3\\ 
F444W && 4.3 & 6.0  & 2.8 & 4.0 & 3.5 & 5.0\\ \bottomrule
\end{tabular}
\label{table: mw max redshift}
\end{table}

\begin{figure*}
    \centering
    \includegraphics[width=\textwidth]{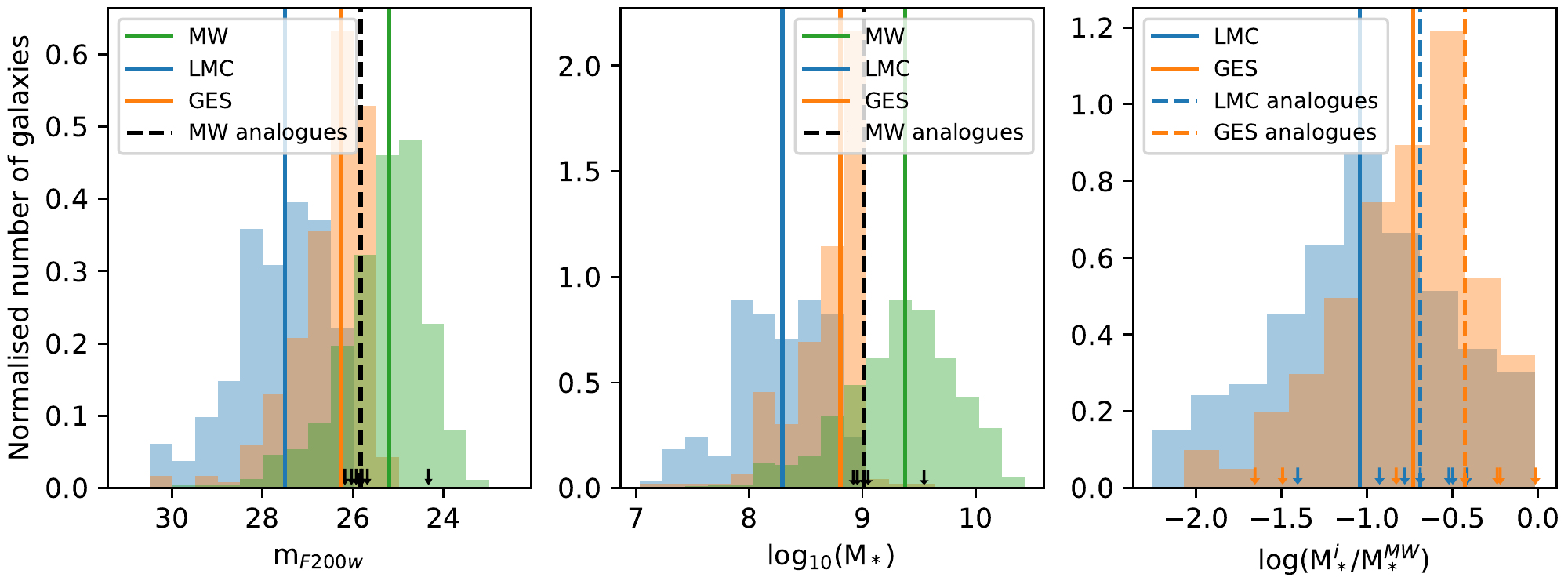}

    \caption{Comparison of the stellar masses and magnitudes of the
      progenitors of MW-, LMC-, and GES-like galaxies at $z=2$.
      \textit{Left}: magnitude distributions, in the F200W passband, for MW-,
      LMC- and GES-like galaxies are shown as green, blue and orange
      histograms, respectively. The medians of the distributions are
      marked with vertical solid lines of similar colour. The small
      black arrows along the x-axis show the magnitudes of the seven
      MW analogues (see text for details) and the vertical black
      dashed line correspond to their median. \textit{Middle}: same as
      the left but for stellar mass distributions.  \textit{Right}:
      the stellar mass ratios of the progenitors of LMC- and GES-like
      galaxies relative to their MW host, shown as blue and orange
      histograms respectively. The solid lines of similar colour mark
      the median of the distributions.  The small arrows along the
      x-axis and vertical dashed lines correspond to the LMC-like and
      GES-like objects associated to the seven MW analogues.}
    \label{fig: histograms mass mag}
\end{figure*}

The first two panels of Fig.~\ref{fig: histograms mass mag} show the magnitude ($m_{\rm F200W}$) and stellar mass distributions of the MW-, LMC-, and GES-like progenitor samples at $z=2$. We can see more clearly here that the progenitors of MW-like galaxies are, on average, brighter and more massive than progenitors of both LMC- and GES-like galaxies. In addition, GES-like galaxies are brighter than LMC-like galaxies with the medians differing by $\Delta(m)\sim 1.2$ and $\Delta(\log_{10}(M_*))\sim 0.5$. The median magnitudes for progenitors of LMC-like galaxies at $z=2$ are consistent with predictions made by \citet[][]{boylan-kolchin_local_2015} who estimate that the LMC would have had a dust-free absolute UV magnitude of -15.6$\pm_{0.6}^{0.8}$. Our dust free absolute M$_{UV}$ for LMC-like galaxies at $z=2$ is M$_{UV}\sim-15.7$. The distribution of masses and magnitudes for MW- and LMC-like progenitor galaxies have a greater spread than GES-like galaxies since the latter were constrained to have a mass between $\Mstar = 0.5-1\times 10^9 \Msun$ around redshift 2 before infall. The third panel of Fig. \ref{fig: histograms mass mag} shows the distribution of the stellar mass ratios between MW-like hosts and each of the LMC- and GES-like progenitor galaxies, all measured at $z=2$. The ratio for GES-galaxies is higher than the ratio for LMC-galaxies by $\Delta(\log_{10}(M^i_*/M^{MW}_*)) \sim0.3$.

In all panels of Fig.~\ref{fig: histograms mass mag}, the dashed
vertical lines represent the median for the MW analogue galaxies, with
individual galaxies shown as small arrows along the x-axis. The left
two panels suggest that the progenitors of MW analogues are more
similar in magnitude and stellar mass to the progenitors of GES-like
galaxies than the MW-like sample as a whole. The right panel shows
that the mass ratios of LMC and GES components of the MW analogue
progenitor systems are higher. This is due to the lower stellar mass of the MW
analogue itself which is lower than the average MW-like galaxy at
higher redshift, as shown by \citet[][]{evans_how_2020}.

\begin{figure*}
    \centering
    \includegraphics[width=\textwidth]{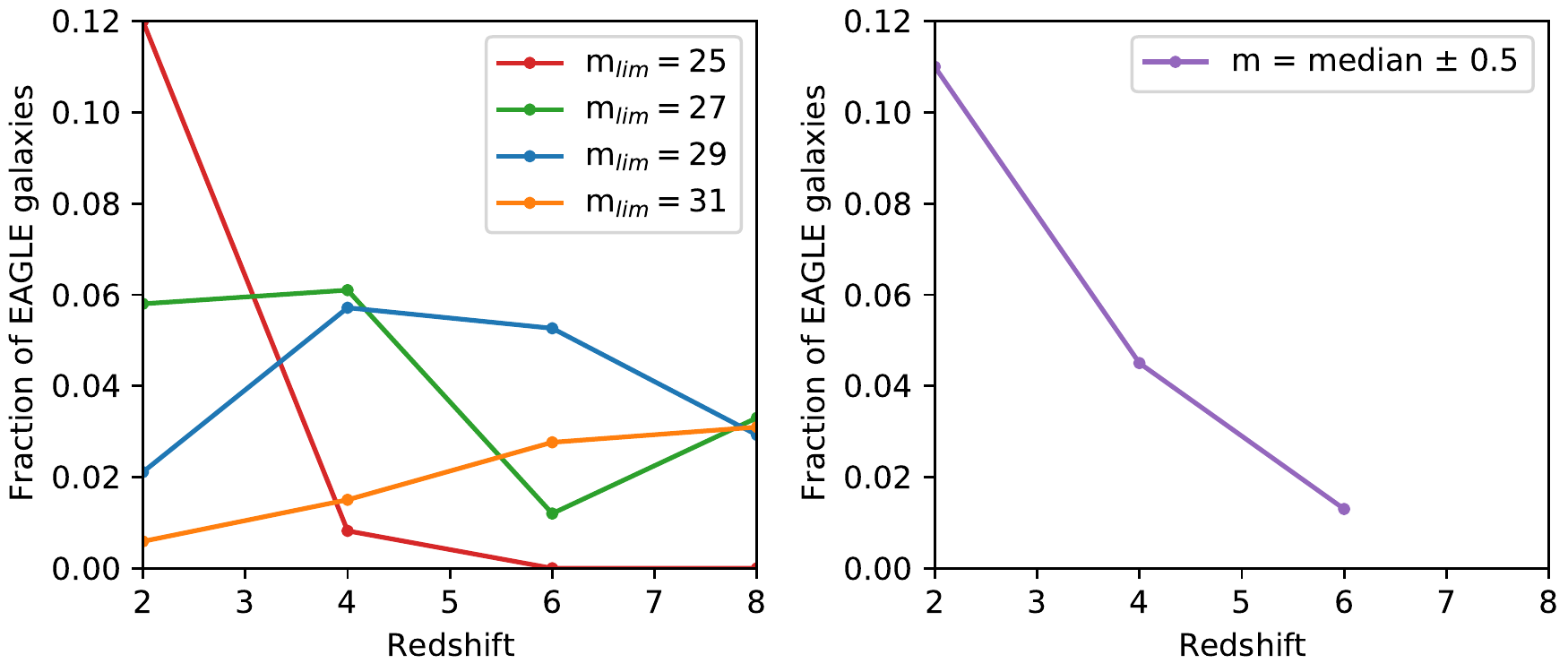}
    \caption{\textbf{Left:} the fraction of EAGLE-Ref galaxies above
      certain limiting magnitudes that are progenitors of MW-like
      galaxies, at different redshifts. The magnitude limits, in the
      F200W \JWST/NIRCam passband, are shown in the legend:
      $m_{\rm lim}=25,27,29,31$ corresponding to red, green, blue and
      orange curves, respectively. \textbf{Right:} the fraction of
      galaxies in EAGLE-Ref that are progenitors of MW-like galaxies,
      and are observable and within a magnitude range corresponding to
      $\pm 0.5$ dex around the median apparent magnitude
      for F200W of Fig.~\ref{fig: mag vs red}, shown in purple.}
    \label{fig: no gals per mw}
\end{figure*}

Not only is it important to know how far back in time the MW
progenitors could be observed, but also to know \textit{how likely}
is it that they will be observed. The left panel of Fig.~\ref{fig: no gals per
  mw} shows the fraction of observable ($m < m_{\rm lim}$) EAGLE-Ref
galaxies that are progenitors of MW-like galaxies, as a function of
redshift. We consider four limiting magnitudes ($m_{F200W}$),
$m_{\rm lim} = 25, 27, 29, 31$, shown in red, green, blue and orange
respectively.  At low redshifts we find $\sim 12 \%$ of galaxies
brighter than $m_{F200W}=25$ to be progenitors of MW-like
galaxies. However, this percentage drops to just $\sim 1 \%$ when
including all galaxies above $m_{F200W}=31$. At high redshift 
($z>6$), there are no longer any galaxies massive/bright enough to
have a magnitude brighter than $m_{F200W}=25$. At $z \sim 8$, the
fainter limiting magnitudes ($m_{\rm lim} = 27, 29, 31$) have the
highest fraction of MW progenitors; $\sim 4\%$ of galaxies are likely
to be progenitors of MW-like galaxies.  

These trends are readily understood. At high redshifts,
  galaxies are less massive and therefore fainter. Thus it is
  extremely unlikely to be as bright as 25 mag. The opposite is true
  for the faintest limiting magnitude ($m_{\rm lim} = 31$) which shows
  an increase in the fraction with redshift. Due to the steep mass
  function the abundances of faint galaxies, at $z=2$, is large and the
  fraction that are MW progenitors is consequently low; by redshift
  $z=8$ those low mass galaxies have dropped below this limiting magnitude
  and the MW progenitors become more prominent.

The fractions of galaxies in each bin shown in the left panel of
Fig.~\ref{fig: no gals per mw} vary considerably with redshift.  In
the right panel of Fig.~\ref{fig: no gals per mw} we use a fixed
magnitude range around the median of MW-like progenitors (shown in the
left panel of Fig.~\ref{fig: mag vs red}).  These ranges correspond to
$\pm 0.5$ dex around the median magnitude for
MW-like progenitors at each redshift, shown in purple. The magnitude range in this panel has its
highest fraction ($\sim 11\%$) at redshift $z=2$ and its lowest
($\sim 1\%$) at redshift $z=6$.  The
  fractions in this panel end at redshift $z=6$ since beyond this time
  more than 50 percent of the progenitors are unidentified (as in
  Fig.~\ref{fig: mag vs red}). At high redshifts ($z=6$) it is clear
that there are many galaxies with a similar magnitude as the MW-like
progenitors that do not become MW-like galaxies by the present.  The
key difference between these galaxies and the progenitors of MW-like
galaxies is simply that they either merge with their host galaxy
(similar to a GES type merger event) or become satellites (similar to
the LMC).

\begin{figure*}
    \centering
    \includegraphics[width=\textwidth]{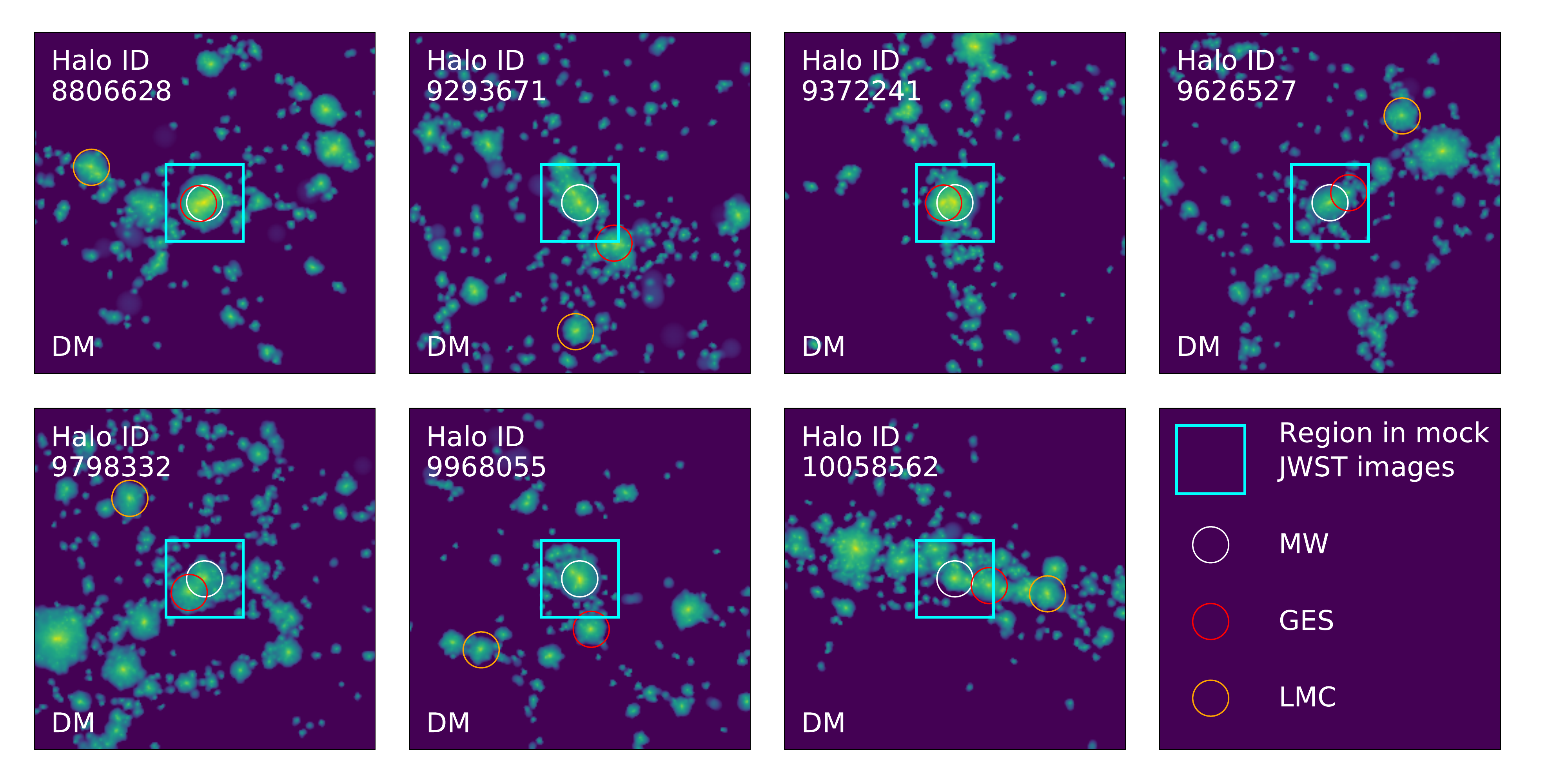}
    
    \caption{Dark matter distribution around seven MW analogues with a
      GES and LMC (see text for details), shown at redshift $z=2$. Each
      panel shows a random projection of particles within a radius of 0.8 Mpc
      centred on the main progenitor of the MW analogue, which is
      marked with a white circle. Red and orange circles represent the
      positions of the centre of the GES and LMC progenitors,
      respectively. Image panels have side length of $\sim$1.13 Mpc, the
      size of the \JWST/NIRCam FoV (2.2 $\times$ 2.2 arcmin) at
      $z=2$. The cyan square in each panel indicates the region size
      for the mock images in Fig. \ref{fig: mock images}. Smoothed
      particle images were made using {\sc py-sphviewer}
      \citep[][]{benitez-llambay_py-sphviewer_2015}.}
    \label{fig: dm all redshift 2}
\end{figure*}

\begin{figure*}
    \centering
    \includegraphics[width=\textwidth]{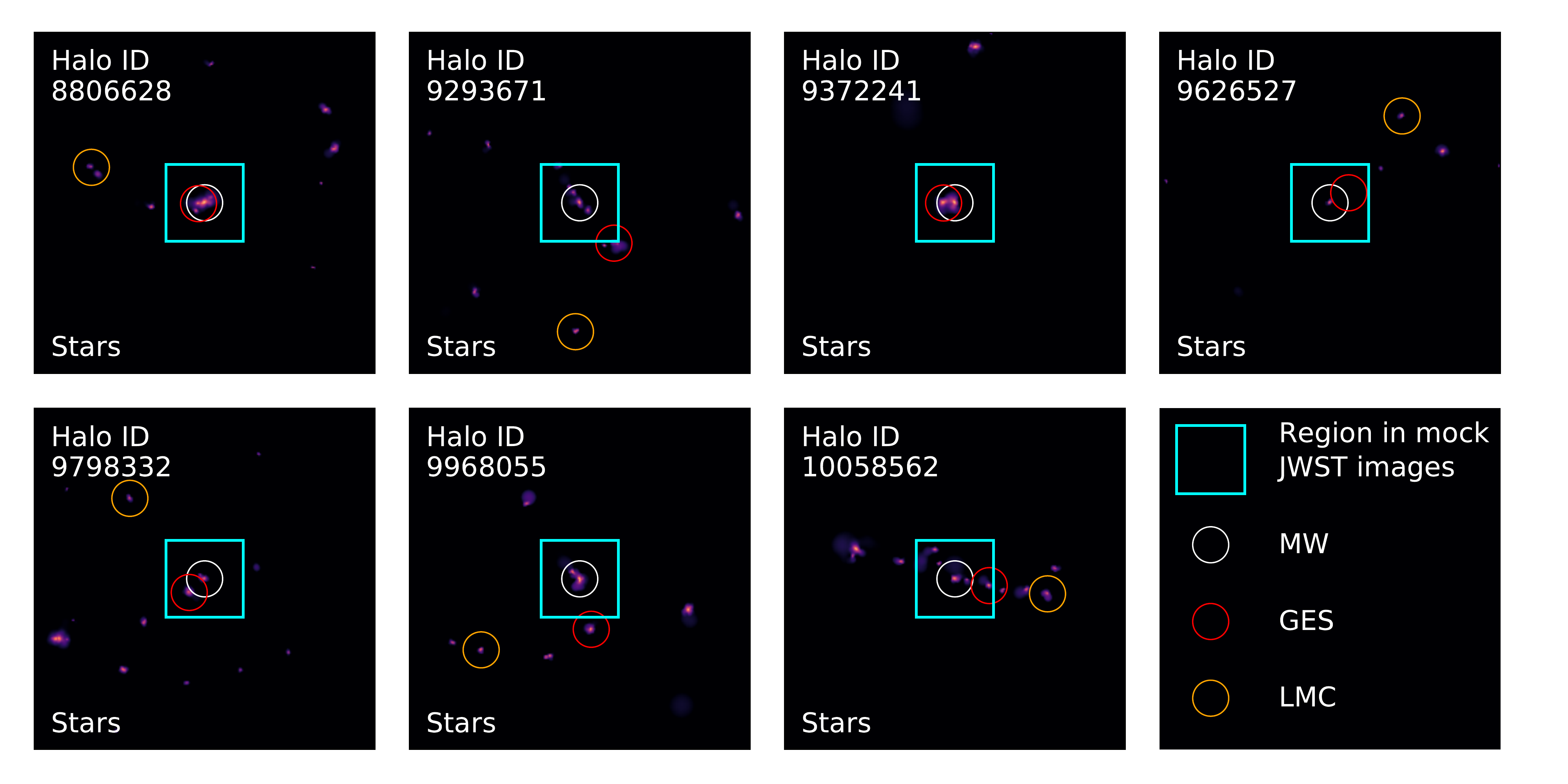}
    
    \caption{Same as Fig.~\ref{fig: dm all redshift 2} but for star particles in the same region.}
    \label{fig: stars all redshift 2}
\end{figure*}

\begin{figure*}
    \centering
    \includegraphics[width=\textwidth]{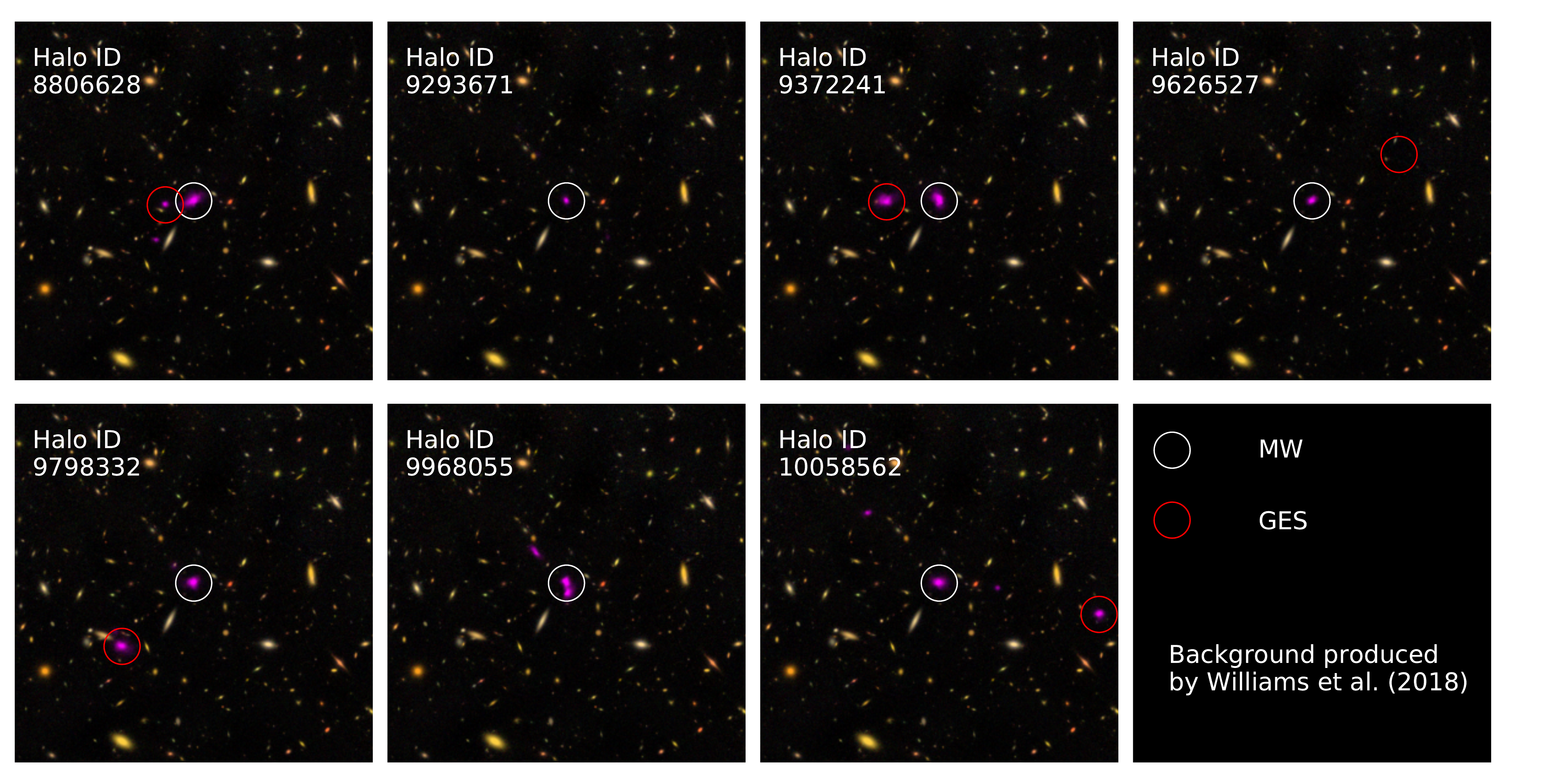}
   
    \caption{Mock \JWST\ images of MW analogue systems, each panel shows a 30 arcsec
      $\times$ 30 arcsec FoV. Stars in the MW analogue systems are shown
      in pink to emphasise their particle locations, this is not
      representative of their observed colour. The background of these
      images were produced by \citet{williams_jwst_2018} to illustrate
      the \JWST\ view of the GOODS-S field. The MW and GES galaxies are
      highlighted using white and red circles, respectively.}
    \label{fig: mock images}
\end{figure*}

\subsection{MW progenitors with realistic accretion histories}

In this section we focus on the small sample of seven MW
analogues with the additional constraints on the accretion
history, namely having a GES-like merger and a LMC satellite. See
Section \ref{sec: methods} and \citet{evans_how_2020} for details. The
dark matter and star particles around these MW analogues at $z=2$ are
shown in Figs.~\ref{fig: dm all redshift 2} and~\ref{fig: stars all
  redshift 2}, respectively. The main progenitor of the MW-like object
is positioned at the centre of each image and is marked with a white
circle. LMC and GES progenitors are also marked with orange and red
circles, respectively. Each panel has a side length of $\sim$1.13 Mpc
which corresponds to the size of the FoV of \JWST/NIRCam (2.2 $\times$
2.2 arcmin) at redshift $z=2$\footnote{Note these are not light cones,
  rather particles at a fixed redshift (fixed snapshot) of the
  simulation.}. These two figures were made using {\sc py-sphviewer} \citep[][]{benitez-llambay_py-sphviewer_2015}, with 64 of the nearest neighbours used for calculating the SPH smoothing length.

GES progenitors are close to the MW progenitors at this
redshift. This is expected as they are constrained to merge with the
main progenitor at $z=1-2$. Interestingly, all of LMC progenitors are
well within the \JWST/NIRCam FoV size. We will elaborate on the distance of
LMC and GES progenitors at various redshifts below.

The cyan squares in Fig.~\ref{fig: dm all redshift 2} and \ref{fig:
  stars all redshift 2} mark the regions of these systems that have
been illustrated in the mock \JWST\ images shown in Fig. \ref{fig: mock
  images}. These have been produced using a background mock image for
\JWST/NIRCam of the GOODS-S field \citep[from][]{williams_jwst_2018} on
to which the images of our MW-analogues have been overlaid. Due to the
small, faint nature of our simulated galaxies, they have been assigned
a pink colour for easy identification in the image. \textit{These
  colours are not illustrative of real life observations.} This figure
shows that without redshift information and potentially other
constraints, identifying the progenitors of the MW and its building
blocks amongst all the foreground and background galaxies will be very
difficult.

\begin{figure*}
    \centering
    \includegraphics{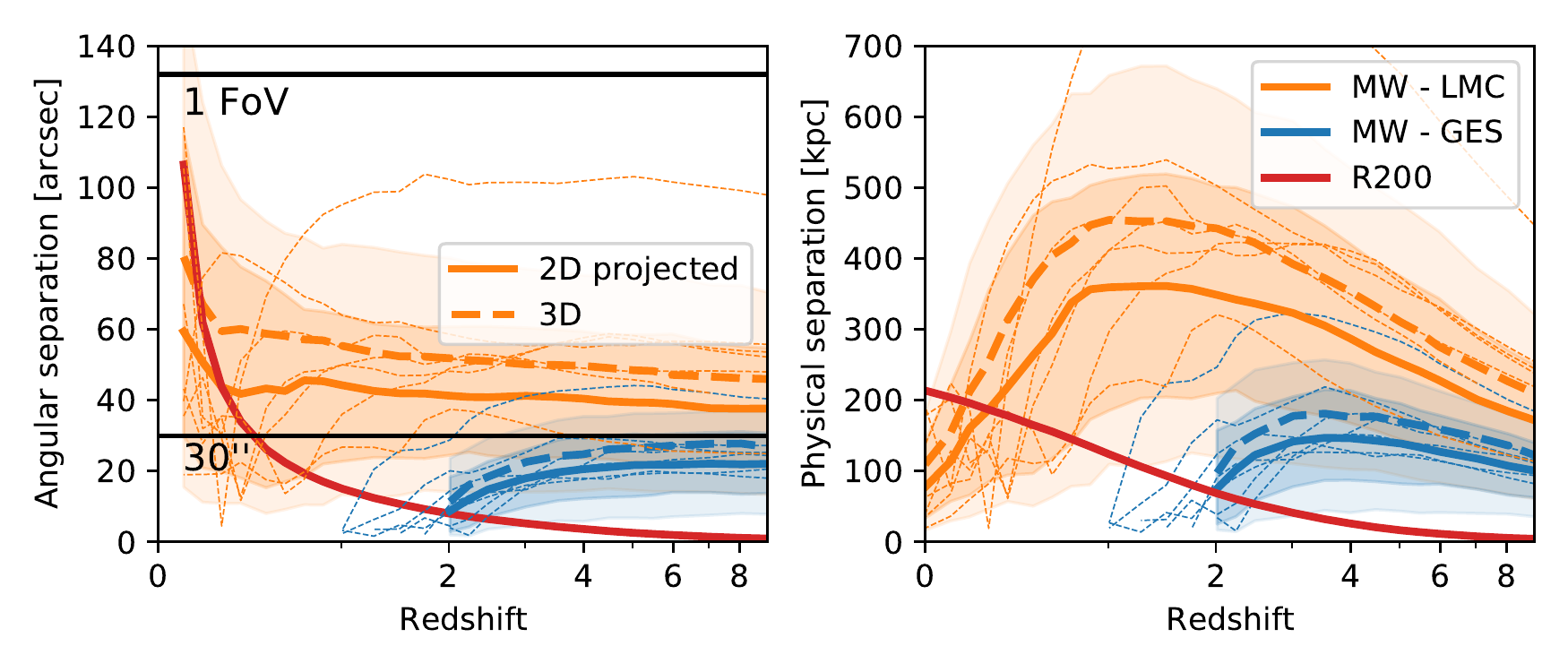}
    \caption{\textit{Left}: Angular separation of LMC-like (orange) and GES-like galaxies (blue) from their host, i.e. the progenitors of MW-like galaxies, as a function of redshift.  
    The thick solid lines and the shaded regions of corresponding colour represent the median, $[16^{th}-84^{th}]$ and $[5^{th}-95^{th}]$ percentile ranges of projected separations for the LMC-like and GES-like groups, whereas the thick dashed lines show the median 3D separations for these groups. Thin dashed lines correspond to 3D separations for the MW analogues group (and their LMC- and GES-like accretions).    
    Since GES is defined as merging between redshifts $z=1-2$, the separations are only shown until redshift $z=2$. Both the LMC- and GES-like galaxies fit within the same \JWST\ FoV as their host, i.e. separation $< 2.2$ arcmin. \textit{Right}: Similar to the left panel but for (proper) physical separation. The red lines in both panels show the median $R_{\rm 200}$ of MW-like galaxies as a function of redshift. The projected separations are based on the average along three orthogonal axes.}
    \label{fig: mw lmc separation}
\end{figure*}

To further investigate the proximity of the LMC and GES progenitors
to the MW progenitor at various redshift, their angular and physical separations are
shown in Fig. \ref{fig: mw lmc separation}. The left panel shows the
median and $[16^{th}-84^{th}]$ percentile of the angular separation
between progenitors of the MW- and LMC-like galaxies, as well as of the MW- and GES-like ones. Angular separations are based on the average of three orthogonal projections. We additionally include individual lines for the subsample of 7 MW
analogues (and the corresponding LMC and GES) where we
show the maximum separation (i.e. 3D distance).

The left panel in Fig.~\ref{fig: mw lmc separation} shows that both
the LMC and GES progenitors fall within the \JWST/NIRCam FoV (120 arcsec)
at all times, with GES progenitors being invariably closer to the MW
than the LMC progenitors. Individual galaxies are shown as fine dashed
lines for the seven MW analogues. Note that one of the LMC satellites
is much further away so is not visible in the ``Halo ID 9372241''
panel in Fig. \ref{fig: dm all redshift 2}. The separations between
MW- and GES-like progenitor galaxies end at redshifts $z=2$ since
this is where some GES galaxies start to merge with their host
galaxies and the median is no longer representative of the whole
sample. The angular resolution limit of \JWST/NIRCam of 0.07 arcsec (at 2
microns) indicates that all GES progenitors can be resolved from
their MW progenitor companion.

The right panel of Fig.~\ref{fig: mw lmc separation} is similar to the
left panel but shows the physical separation. The turnaround time and
infall time of the objects are easier to see here. The
$R_{\rm 200}$ evolution of a MW
analogue is shown with a red curve in both panels for
reference. LMC-like satellites have a recent infall time, $z\sim 0.3$,
consistent with previous works
\citep[e.g.][]{boylan-kolchin_dynamics_2011,rocha_infall_2012}. Such
massive satellites are affected by dynamical friction to a large
degree and they merge quickly with the host; hence those surviving at
redshift at $z=0$ must have fallen recently
\citep[e.g][]{fattahi_tale_2020}.  The turnaround redshift and radius
of the LMC sample are on average $z\sim1.5$ and $r=360\pm160$kpc,
respectively. GES analogues have a smaller turnaround radius ($r=150\pm60$kpc) and
earlier accretion times ($z\sim3.5$), compared to LMC progenitors. This is expected since GES are constrained to merge with the MW progenitors by $z=1$.

Combining the results from Fig.~\ref{fig: mag vs red} and
Table~\ref{table: mw max redshift}, we conclude that MW-like
progenitor galaxies should be observable up until $z\sim6$ in most
\JWST/NIRCam passbands, with associated LMC- and GES-like galaxies observable
until redshifts $z \gtrsim 4$ and 5.3 respectively. At these times the
LMC- and GES-like galaxies will most likely be within the \JWST/NIRCam FoV.

\subsubsection{Colour-magnitude diagrams}
\begin{figure*}
    \centering
    \includegraphics{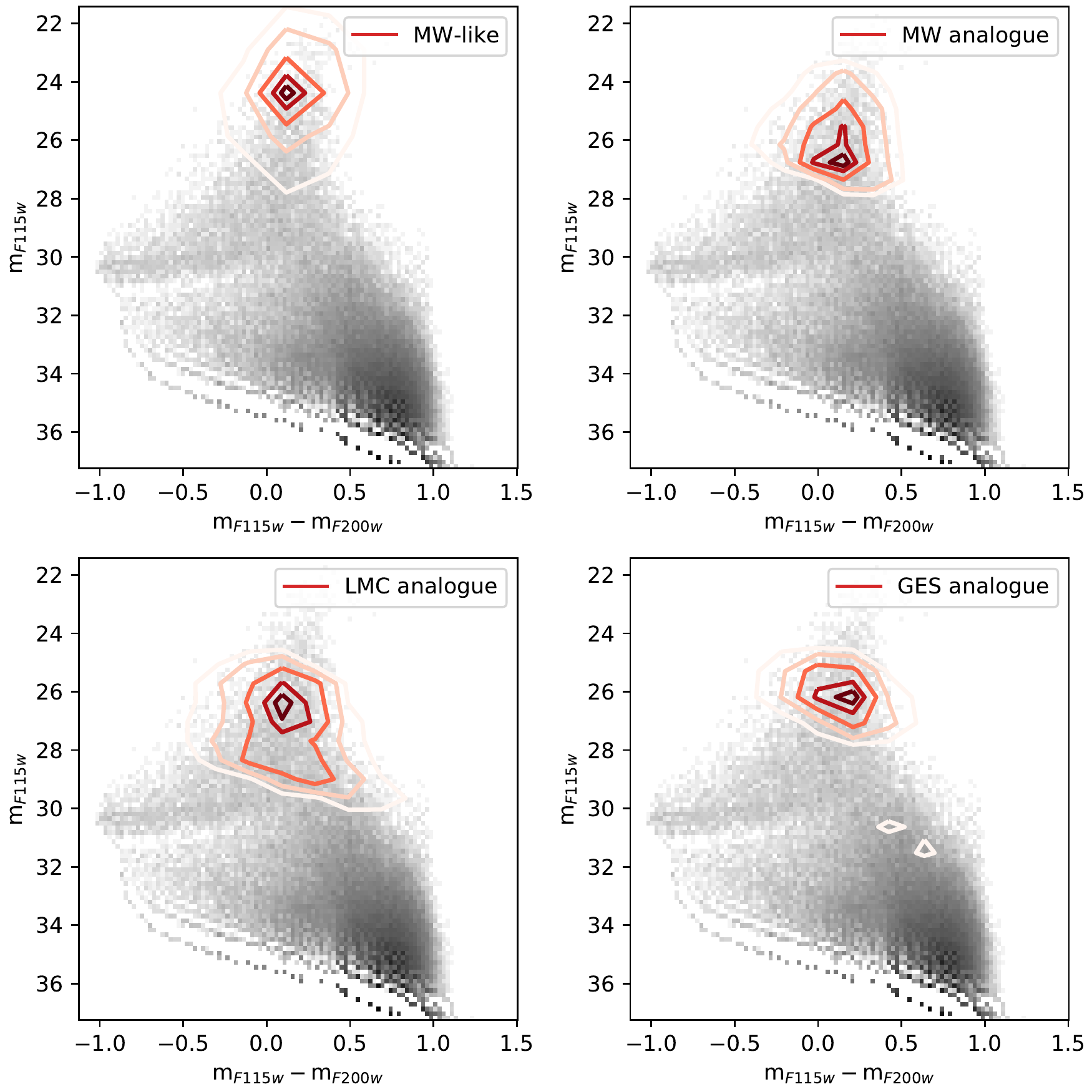}
    
    \caption{The CMDs for galaxies in the F115W and F200W
      passbands at redshift $z=2$. The grey-scale background shows the
      colour-magnitude distribution for all galaxies in EAGLE-Recal;
      the red contours represent the area of the colour-magnitude
      diagram that 10, 25, 50, 75 and 90 percent of the progenitors of MW-like, MW analogues and their
      LMC and GES components would be contained in the upper left to
      lower right panels, respectively. The mass distribution of
      MW-like galaxies is quite broad, which causes a large spread in
      magnitudes.}
    \label{fig: cmd z2}
\end{figure*}

We now consider whether or not progenitors of the MW, LMC and GES are
distinguishable from other galaxies at the same redshift. We turn to
the colour-magnitude diagrams (CMDs) for another layer of
information. Fig. \ref{fig: cmd z2} shows the CMD for the progenitor
in the F115W vs. F115W-F200W plane at $z=2$. These passbands were chosen
at random since at $z=2$ there was minimal differences between CMDs.

We show all MW-like galaxies, and the individual components of the
seven MW analogues (MW, LMC and GES) in various panels in
Fig.~\ref{fig: cmd z2} from top left to lower right, respectively. The
background shows a grey-scale density distribution for the overall
colour-magnitude distribution of the total population of galaxies in
the EAGLE-Recal run which has higher resolution. At fainter magnitudes
we see individual `ridges' that are likely caused by resolution
effects. Similarly to the approach used for Fig.~\ref{fig: mag vs
  red}, we do not use the magnitudes (and colours) directly from the
EAGLE-Ref run for our target galaxies, especially because the LMC and
GES progenitors at high redshifts have relatively low stellar
mass. Instead, we highlight with red contours the location where
galaxies with similar stellar masses to our target galaxies lie on the
CMD.

The top left panel of Fig.~\ref{fig: cmd z2} shows that the
progenitors of MW-like galaxies are among the brightest galaxies at
$z=2$ with magnitudes brighter than $m_{\rm F115W} \sim 26$. However,
MW analogues (top-right panel) lie at the lower magnitude ranges for
all MW-like galaxies. This is consistent with the stellar mass
evolution shown in \citet{evans_how_2020} where MW analogues have a
much lower mass than typical MW-like galaxies. The MW, LMC and GES
analogues all lie within a similar space in the colour-magnitude
diagram, with a greater range in colour space than magnitude
space. They tend to have magnitudes around 26 and colour between -0.5
and 0.5, in these passbands.

We include a similar figure but for CMDs at redshift $z=4,6$ and 8 in
Appendix \ref{sec: high z cmd}. In summary, the results discussed
above hold at those redshifts too. We note that at $z=6$ and 8, some
combination of colours in \JWST/NIRCam passbands will not yield useful CMDs, as
some passbands are in the Lyman-break, as shown in Appendix ~\ref{sec: spectra}. More precisely, F070W at
$z=6$ is bluer than the Lyman-break, and both F070W and F090W are
bluer by $z=8$.

\section{Discussion and Conclusions} \label{sec: disc and conc}

We provide predictions for \JWST\ using cosmological hydrodynamical simulations from the EAGLE project \citet{schaye_eagle_2015}. We have calculated dust-free magnitudes in \JWST/NIRCam bands for all galaxies, and then applied a simple-analytic dust correction using the ISM column density and temperature along the line-of-sight \citep[based a modified version of Dust Model B from][]{vogelsberger_high-redshift_2020}. The dust-corrected
magnitudes were used to produce comoving galaxy luminosity functions
at redshifts $z=2-8$ along with the estimated number of galaxies in a
\JWST/NIRCam FoV across the same redshift range. In the second half of
this paper we focused on MW analogues and the main accreted objects onto them, namely LMC- and GES-like objects, as identified in
\citet{evans_how_2020}, to see how far back in time their progenitors
might be observable, and if it might be possible to identify them in a
\JWST/NIRCam FoV. Our main conclusions are as follows:

\begin{itemize}
\item We compare our results with those from \citet{vogelsberger_high-redshift_2020} which is based on Illustris-TNG, and a more sophisticated treatment of dust attenuation using radiative transfer code, SKIRT. We find excellent agreement between the two results when comparing $\Mstar$ vs. dust-corrected magnitudes ($M_{\rm UV}$).

\item Our luminosity functions are in overall agreement with those from \citet{vogelsberger_high-redshift_2020}. Considering the previous point, the differences in the luminosity functions are likely caused by differences in the stellar mass functions of EAGLE and Illustris-TNG, resulting from their different subgrid galaxy formation models. Our luminosity functions are also in good agreement with those produced from the JAGUAR mock catalogue for \JWST \citep[][]{williams_jwst_2018}, however, the JAGUAR luminosity functions are flatter at the bright end and hence our results may underestimate the number of bright galaxies observable with \JWST/NIRCam.

\item The best-fit parameters for the Schechter functions were used to
  predict the expected number counts of galaxies at each redshift. We
  expect a maximum of $\sim2400$ galaxies at redshift $z=2$ and
  $\sim 80$ at redshift $z=8$ for an exposure time of 10$^5$s ($SNR=5$). These numbers reduce to $\sim1300$ and $\sim 8$ at those two redshifts, respectively, for exposure time of  10$^4$s ($SNR=10$). These predictions are overall lower than the average numbers from \citet{vogelsberger_high-redshift_2020}. This discrepancy does not affect the MW progenitor results because MW progenitor galaxies are among the fainter galaxy population. It would, however, affect the numbers predicted in a field of view, especially at redshift 8 since at this redshift, the counts are no longer dominated by the faint end for the $10^4$s exposure. We found that our predicted numbers of galaxies are consistent with \citet[][]{cowley_predictions_2018}, who used the semi-analytic model of galaxy formation, Galform.

\item Assuming an exposure time of 10$^5$s and $SNR=5$, a MW-like
  progenitor galaxy would be observable with \JWST\ up to redshift
  $z \sim 6$, whereas progenitors of LMC- and GES-like galaxies would
  be observable out to redshifts $z\sim4$ and 5.3, respectively. The
  optimal passband is F356W and the least sensitive is F070W. In the
  F356W passband, \JWST\ should be able to observe galaxies on average out
  to $\Delta z = 1$ more than in the F070W passband. These limits reflect
  the fact the these passbands have the best and worst transparency
  respectively.

\item The progenitors of the individual components of the MW analogue systems (MW, LMC, GES galaxies) have very similar stellar masses and magnitudes at high redshifts, with GES analogues being on average slightly more massive than the LMC analogues. The main difference in their fate lies in whether they become a host galaxy, satellite galaxy or if they merge with their host galaxy.

\item The median magnitudes of progenitors of LMC-like galaxies at $z=2$ is consistent with predictions made by \citet[][]{boylan-kolchin_local_2015} who estimate that the LMC would have had a dust-free absolute UV magnitude of -15.6$\pm_{0.6}^{0.8}$. Our dust free absolute M$_{UV}$ for LMC-like galaxies at $z=2$ is M$_{UV}\sim-15.7$.

\item Our results suggest that the progenitors of the LMC- and GES-like galaxies always lie within 60 and 30 arcsec, respectively, of MW progenitors at all times and therefore will fit within one FoV of \JWST/NIRCam.

\item The CMDs of the progenitors of MW
  analogues also suggest that the three components (MW, LMC, GES)
  should lie in a similar colour-magnitude range. Galaxies of similar
  mass to the MW-, LMC- and GES-like galaxies in the MW analogue
  systems have a wide range of colours but a narrow range in 
  magnitude.

\end{itemize}

In summary, our simulations indicate that it should be possible to
observe progenitors of MW analogues using \JWST\ and also observe the
progenitors of their LMC-like satellites and GES-like companions at
early times. Up until the redshift at which they are observable
(typically $z=4$), the three galaxies should all fall within the same
FoV. At redshift $z=2$ galaxies with similar mass and $m_{\rm F115W} \sim 26$ could be analogues to the MW/GES merger.  
This is an exciting opportunity to link the high
redshift universe to our galaxy today.

In closing, we remark that our study can be extended and refined with future generations of simulations, which will provide larger volumes and/or finer resolution. 
The EAGLE-Ref simulations span 100Mpc$^3$ but, there are only 7 MW-analogues in this volume. Larger simulations will allow for better statistics and hence firmer conclusions can be made for MW analogue galaxies. With better statistics we could also investigate MW-like systems within the Local Group environment \citep[and hence provide comparisons with work such as][]{boylan-kolchin_local_2016,santistevan_formation_2020}. Finally, higher resolution simulations would allow us to calculate the surface brightness and size of low mass galaxies \citep[which would allow for comparisons with work such as][]{patej_detectability_2015}.



\section*{Acknowledgements}

We thank Calvin Sykes for sharing the ENGinE data with us, we also thank Mathilde Jauzac and Renkse Smit for their helpful comments on this manuscript. AD thanks the staff at the Durham University Day Nursery who play a
key role in enabling research like this to happen.  TE is supported by
a Royal Society Research Grant.  AD is supported by a Royal Society
University Research Fellowship. AD acknowledges support from the
Leverhulme Trust and the Science and Technology Facilities Council
(STFC) [grant numbers ST/P000541/1, ST/T000244/1]. AF is supported by
the UK Research and Innovation (UKRI) Future Leaders Fellowships
(grant numbers MR/V023381/1, MR/T042362/1).  CSF acknowledges European
Research Council (ERC) Advanced Investigator grant DMIDAS (GA
786910). This work used the DiRAC@Durham facility managed by the
Institute for Computational Cosmology on behalf of the STFC DiRAC HPC
Facility (www.dirac.ac.uk). The equipment was funded by BEIS capital
funding via STFC capital grants ST/K00042X/1, ST/P002293/1,
ST/R002371/1 and ST/S002502/1, Durham University and STFC operations
grant ST/R000832/1.  DiRAC is part of the National e-Infrastructure.

\section*{Data Availability}
The data used in this article are available in the EAGLE online
database, at \url{http://virgodb.dur.ac.uk:8080/Eagle}.




\bibliographystyle{mnras}
\bibliography{references} 



\appendix

\section{Resolution checks and magnitude corrections}\label{sec: app mag correct}
\begin{figure}
    \centering
    \includegraphics[width=\columnwidth]{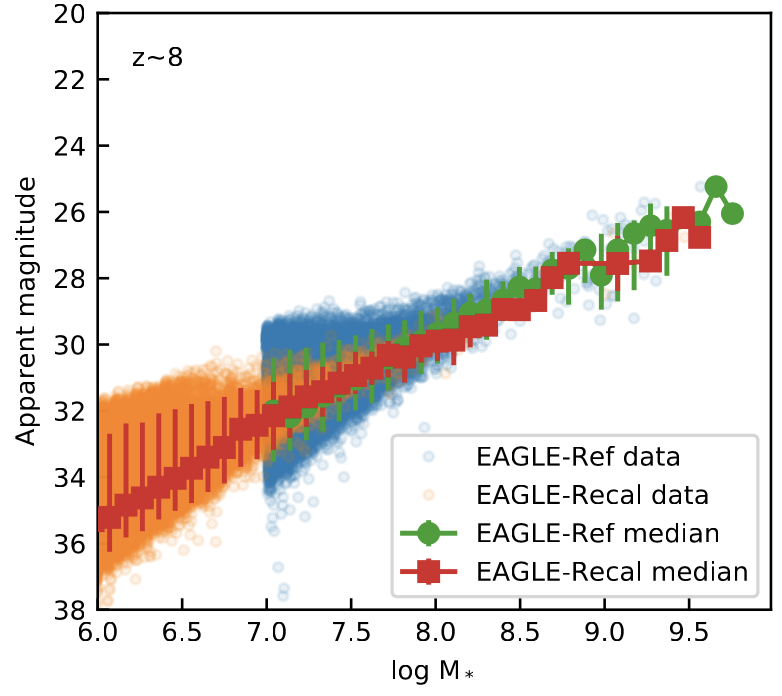}
    \caption{The relationship between apparent magnitude and stellar
      mass for the EAGLE-Ref (blue points) and EAGLE-Recal (orange)
      volumes at $z=8$. The median values for each sample are shown as
      green circles and red squares for EAGLE-Ref and EAGLE-Recal,
      respectively. The errorbars represent the $[5^{th}-95^{th}]$
      percentiles of the data.}
    \label{fig: resolution checks}
\end{figure} 

We compare the dust-corrected apparent magnitudes for all
galaxies in the EAGLE-Ref and EAGLE-Recal volumes in order to quantify
the effects of resolution on our results. The relationship between
apparent magnitude and stellar mass for the two volumes is shown in
Fig.~\ref{fig: resolution checks}. The median values for the two
volumes are consistent with each other. However, as shown in the
figure, there is a large amount of scatter at the low mass end for the
EAGLE-Ref volume due to the lack of resolution.

In order to get a better representation of the magnitudes at the low
mass end, we apply a magnitude correction to galaxies of stellar mass
$M_* < 10^8 \Msun$ as this is where the scatter in the EAGLE-Ref data
starts to increase dramatically.  Such corrections, however, are not
necessary for brighter galaxies as these are resolved with enough star
particles.  (See agreement at the bright end in Fig.~\ref{fig:
  resolution checks}).


For the magnitude corrections below $M_* < 10^8 \Msun$, we model the
distribution of apparent magnitudes for the higher resolution Recal50
dataset in a given mass bin between the $[5^{th}-95^{th}]$ percentile
values (shown as red errorbars on Fig.~\ref{fig: resolution checks})
and resample values for EAGLE-Ref from this distribution. These
corrections are performed at all redshifts for magnitudes with and
without dust, and for all passbands used in this work.

\begin{figure}
    \centering
    \includegraphics[width=\columnwidth]{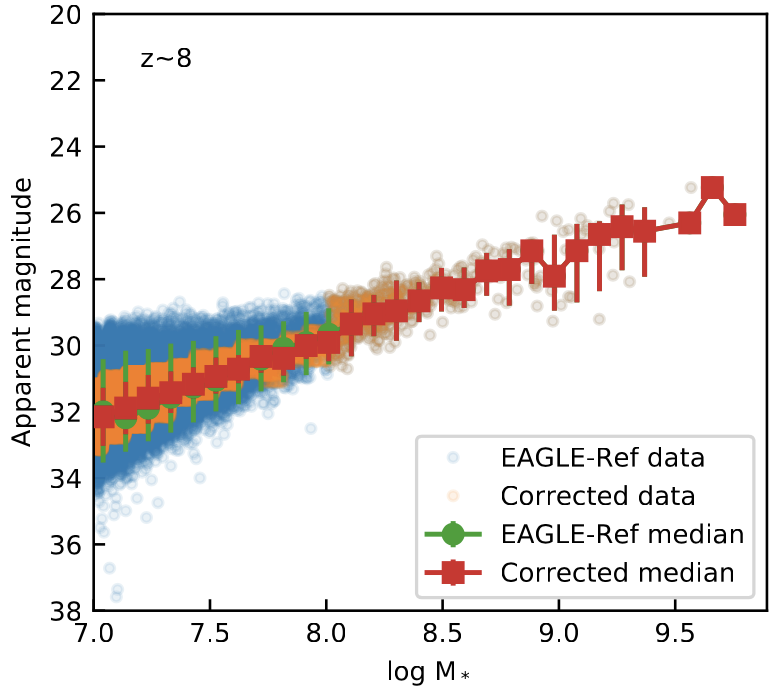}
    \caption{The apparent magnitude-stellar mass relation for the
      uncorrected (blue points) and corrected (orange points)
      EAGLE-Ref data.} 
    \label{fig: corrections new vs old}
\end{figure}

Fig. \ref{fig: corrections new vs old} shows the original data from 
EAGLE-Ref (blue points) and the new corrected magnitudes (orange
points). As shown in the figure, the median values are unchanged. The
corrected magnitude data points no longer show such a large scatter at
the low mass end.

\section{CMDs at high redshift}\label{sec: high z cmd}

\begin{figure*}
    \centering
    \includegraphics[width=0.4\textwidth]{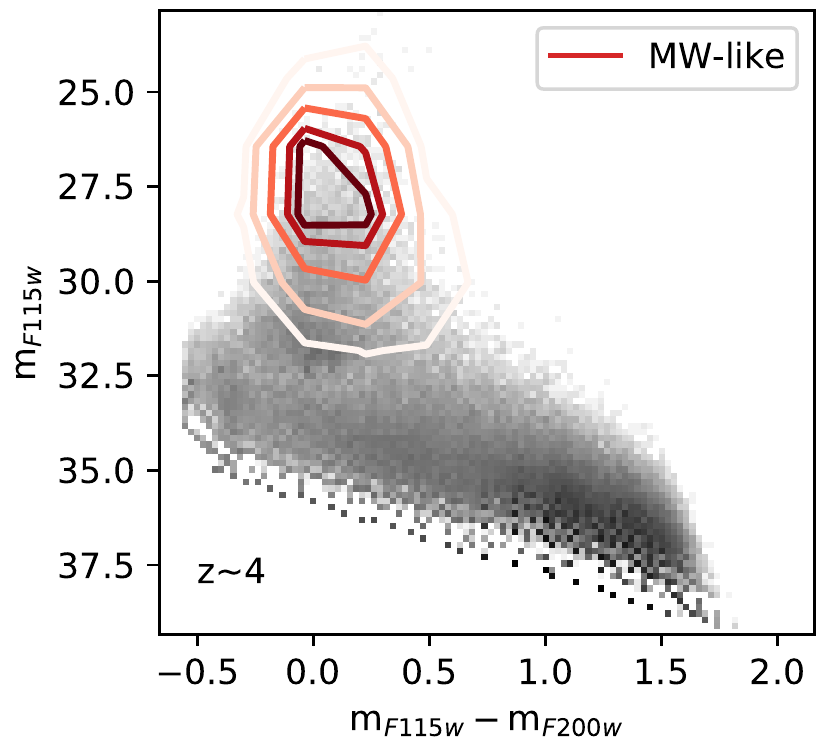}
    \includegraphics[width=0.4\textwidth]{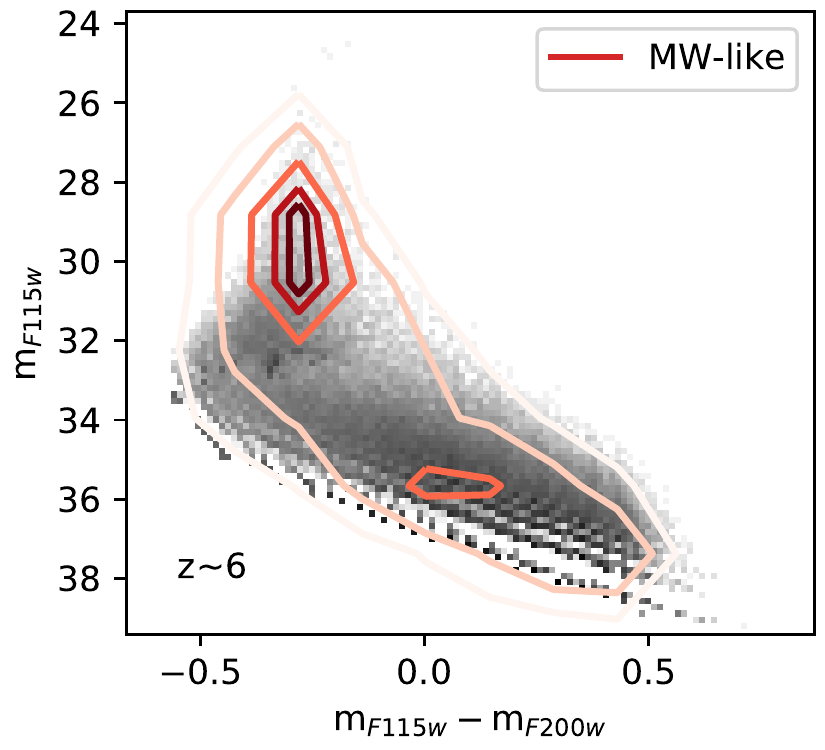}
    \includegraphics[width=0.4\textwidth]{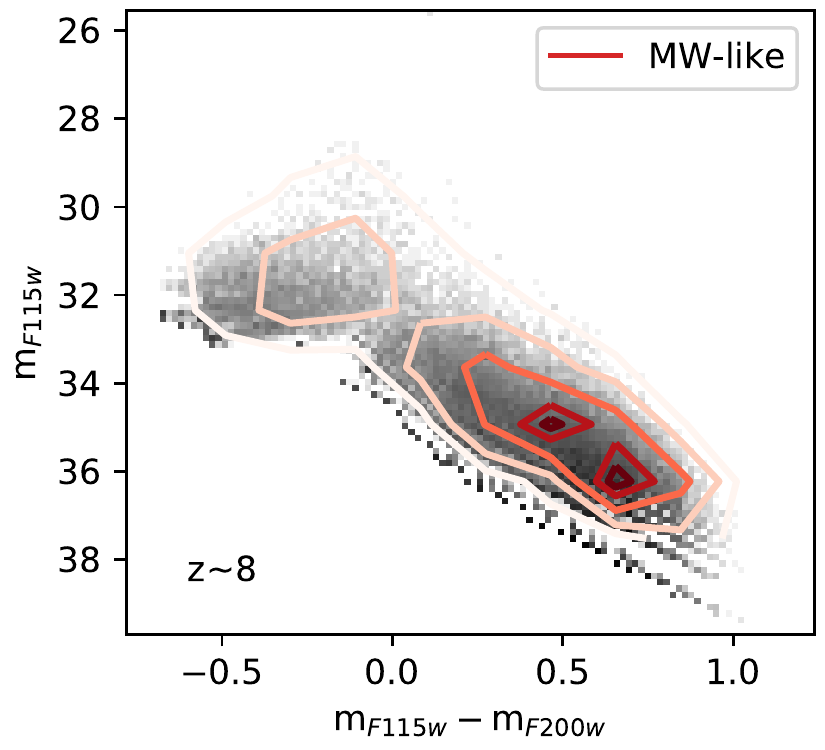}
    \caption{CMDs for all EAGLE-Recal galaxies (grey) at redshifts $z=4,6,8$ with overlaid red contours to show  the region where 10, 25, 50, 75 and 90 percent of MW-like galaxies are enclosed.}
    \label{fig: cmd high redshifts}
\end{figure*}

Fig.~\ref{fig: cmd high redshifts} shows the CMDs
for the $m_{\rm F115W}$ and $m_{\rm F200W}$ passbands at high
redshifts. The red contours in each redshift panel highlight the
regions where progenitors of MW-like galaxies are most likely to
reside. As the redshift increases, the MW-like
progenitors occupy a larger region of the CMD; by redshift $z=8$, the
contours cover the entirety of the CMD, but are
more concentrated at fainter magnitudes and redder colours.

\section{Milky Way analogues as Lyman break galaxies}\label{sec: spectra}

\begin{figure*}
    \centering
    \includegraphics{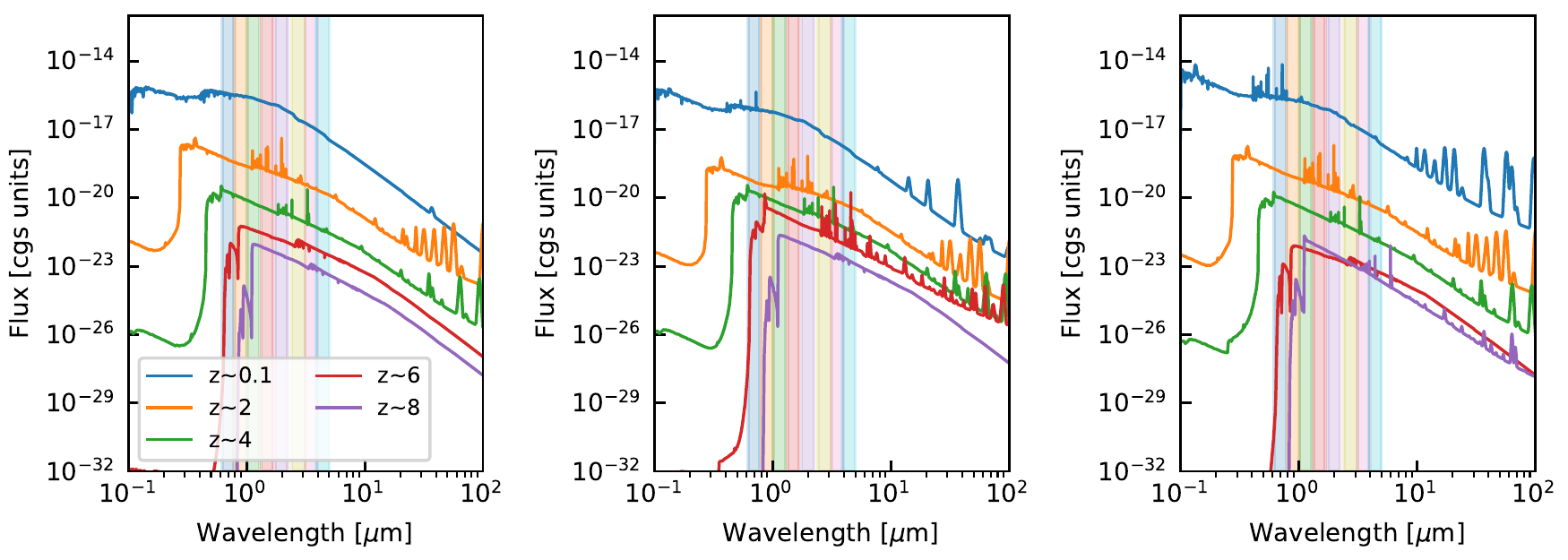}
    \caption{The dust free spectra of the progenitors of three of the
      MW analogues. Spectra are shown for each galaxy at redshifts
      $z=0.1, 2, 4, 6, 8$ in blue, orange, green, red, and purple
      respectively. The vertical coloured bands represent the
      \JWST/NIRCam passbands (from left to right: F070W, F090W, F115W,
      F150W, F200W, F277W, F356W, F444W). MW analogues may have Lyman
      break features at high redshifts, i.e. $z=6$ and 8.}
    \label{fig: spectra mw}
\end{figure*}

The dust-free spectra of three MW-analogues are shown in
Fig.~\ref{fig: spectra mw}. These spectra were made using the MILES
Stellar library as part of FSPS, which include features such as
absorption from the intergalactic medium and emission from nebulae.

The higher redshift spectra for the progenitors of these galaxies are
shown as red and purple lines for $z=6, 8$ respectively. These indicate
that at these high redshifts, the progenitors of MW analogues could be
seen as Lyman break galaxies in the bluer wavelength passbands,
i.e. F070w, F090w and F115W. The Lyman break features of these spectra
may help distinguish MW-analogues from other galaxies in the field.


\bsp	
\label{lastpage}
\end{document}